\documentclass[superscriptaddress,floatfix,letterpaper,accepted=2023-02-24]{quantumarticle}
\pdfoutput=1

\usepackage{float}
\usepackage{graphicx}  
\usepackage{xcolor}
\usepackage{bm}        
\usepackage{braket}
\usepackage{amssymb}   
\usepackage{epstopdf}
\usepackage{xfrac}
\usepackage{cancel}
\usepackage{soul}
\usepackage{amsmath}
\usepackage{amsthm}
\usepackage{amsfonts}
\usepackage{bbm}
\definecolor{darkblue}{rgb}{0.1,0.2,0.6}
\definecolor{darkred}{rgb}{0.8,0.1,0.2}
\definecolor{darkgreen}{rgb}{0.31,0.62,0.24}
\definecolor{quantumviolet}{rgb}{0.325,0.145,0.498}
\usepackage[colorlinks,citecolor=darkblue,linkcolor=darkblue,urlcolor=quantumviolet]{hyperref} 
\usepackage{enumerate}
\usepackage{setspace}
\usepackage{url}  
\usepackage{mathrsfs}
\usepackage{mathtools}

\newcommand{\cnot}[1]{\textsc{cnot}_{#1}}
\newcommand{\cz}[1]{\textsc{cz}_{#1}}

\newcommand{\vac}{\ket{\text{vac}}}

\numberwithin{equation}{section}
\renewcommand\theequation{\arabic{section}.\arabic{equation}}

\usepackage[numbers,sort&compress]{natbib}
\begin{document}

\title{Modular architectures to deterministically generate graph states}

\author{Hassan Shapourian}
\affiliation{Cisco Quantum Lab, San Jose, CA 95134, USA}
\author{Alireza Shabani}
\affiliation{Cisco Quantum Lab, Los Angeles, CA 90049, USA}


\begin{abstract}
Graph states are a family of stabilizer states which can be tailored towards various applications in photonic quantum computing and quantum communication. 
In this paper, we present a modular design based on quantum dot emitters coupled to a waveguide and optical fiber delay lines  to deterministically generate N-dimensional cluster states and other useful graph states such as tree states and repeater states. Unlike previous proposals, our design requires no two-qubit gates on quantum dots and at most one optical switch, thereby, minimizing challenges usually posed by these requirements. Furthermore, we discuss the error model for our design and demonstrate a fault-tolerant quantum memory with an error threshold of $0.53\%$ in the case of a 3d graph state on a Raussendorf-Harrington-Goyal (RHG) lattice. We also provide a fundamental upper bound on the correctable loss in the fault-tolerant RHG state based on the percolation theory, which is $1.24$ dB or $0.24$ dB depending on whether the state is directly generated or obtained from a simple cubic cluster state, respectively.
\end{abstract}

\maketitle

\section{Introduction}

Photonic systems offer a promising route for scalable quantum computing and networking (collectively called quantum information processing). 
Integrated photonic chips have a small footprint and are network compatible in the sense that they can be connected together via an optical quantum network to realize distributed quantum computing~\cite{OBrien2009,Bogdanov:17}.
Another feature of photon-based qubits compared to matter-based qubits is that photonic qubits are protected from decoherence since they barely interact with environment. However, developing quantum photonic technologies poses two new challenges: First, photons do not easily interact with each other and require nonlinear devices to mediate interactions\footnote{There is a separate line of research on quantum computing~\cite{Knill2001} and realizing gates based on linear optics which led to development of probabilistic postselected and fusion gates~\cite{PhysRevA.65.062324,PhysRevA.66.024308,Browne2005}. We do not address these approaches in this work.}.
Even though integrating several devices in the form of a quantum circuit (where photons interact by flowing through) can be done for a specific task, it is not a general purpose solution.
Second, photonic systems suffer from loss error where photons either are absorbed or leak out.


Fortunately, both issues can be addressed in a measurement-based scheme~\cite{Briegel2009,Zwerger2016} where various tasks including applying quantum gates, error (loss) correction, teleportation, etc.~can be 
realized in terms of a sequence of (easily reconfigurable) single-qubit measurements on a multi-photon entangled state, a.k.a.~resource state. 
 Resource states are usually realized in the form of a graph state.
 Among others, two important advantages of using graph states are that they are stabilizer states 
 which means the stabilizer formalism toolbox is available
 and they are equipped with a graph representation which provides room for further innovations by applying graph theory ideas.
 In the context of computing,
cluster states in two and three dimensions are archetypal examples of graph states which can be used to perform universal~\cite{Raussendorf-2d,PhysRevLett.86.5188,Nielsen2004} and fault-tolerant~\cite{Raussendorf-ftcs} quantum computation, respectively. In the context of networking, quantum repeater states and
loss-tolerant quantum error correcting codes, based on all-to-all connected~\cite{PhysRevA.85.062326,PhysRevLett.110.260503,Azuma2015}, self-similar~\cite{PhysRevA.94.052307}, and tree graphs~\cite{Borregaard2020,Morley_Short_2019}, respectively,
have been developed as new generations of quantum communication protocols. Consequently, design and implementation of photonic devices to efficiently generate resource graph states play a crucial role in realizing measurement-based photonic quantum information processing. Our focus in this work is on optimal designs for a deterministic resource state generator.


Different protocols to generate entangled resource states generally fall into two categories~\cite{Orieux2017,Moody2022}: One involves probabilistic sources of entangled photon pairs, based on three-wave or four-wave mixing, whose output is fed into a linear optical circuit for further probabilistic gating and postselection of the desired graph state~\cite{Adcock2019}. Despite the technological breakthroughs in this field, the probabilistic nature of this approach causes a large overhead and has limited its scalability. The other approach is based on light-matter interfaces or quantum emitters which are deterministic sources of single photons~\cite{Aharonovich2016}. 
A semiconductor quantum dot (QD)~\cite{Senellart2017,Javadi2018} in a nanophotonic cavity or waveguide
is a prototypical example of a quantum emitter which can further be used as a source of nonlinearity~\cite{Jeannic2021,Schrinski2021} and integrated with photonic chips.
Quantum emitters have also been realized in other systems such as atoms, ions, vacancy centers, and molecules. The main advantage of the latter approach is that the photon source and nonlinear gates are deterministic, which leads to high resource-state generation rates without a large overhead. Additionally, recent advances in QD technologies \cite{Uppu2020,Tomm2021,Uppu2021}
 offer a great deal of quantum controllability and makes it possible to achieve large Purcell enhancement and produce indistinguishable photons over extended timescales.

Designing optical circuits and protocols for deterministic generation of entangled states of photons such as Greenberger–Horne–Zeilinger (GHZ) and linear cluster states via a single quantum emitter was pioneered in Ref.~\cite{Lindner-Rudolph2009}, which was subsequently demonstrated experimentally using a QD~\cite{Schwartz2016} and more recently using a single atom in a cavity~\cite{Thomas2022}.
The core idea is that by applying a temporal sequence of resonant pulses to a quantum emitter whose logical state is controlled by a separate optical transition, we obtain a sequence of entangled photons at the output.
This scheme was extended to coupled QDs for generating 2d cluster states~\cite{PhysRevLett.105.093601,PhysRevLett.123.070501}, and repeater states~\cite{Economou2017},
which was followed up and further analyzed in several recent studies~\cite{Russo_2019,Hilaire2021resource,Li2021}.
Despite being theoretically interesting, experimentally realizing such circuit designs are often extremely challenging since they require multiple actively monitored emitters and  two-qubit gates on them. Even if the two-qubit gates were feasible, it would limit the overall generation rate of resource states as these gates are typically slow. Furthermore, unlike the original proposal, these designs typically lack a regular repeated pattern of resonant pulses and emitter gates, which leads to an irregular temporal output. In other words, there is no obvious way to define a reference clock cycle for outgoing photons. This is particularly an important issue for quantum communication purposes where clock synchronization is vital.

In a parallel direction, Ref.~\cite{PhysRevLett.116.093601} opened another route by designing a novel resource state generator in terms of a quantum emitter coupled to a quantum feedback loop (or a time-delay line), which was further pursued to design protocols for generating cluster states in 2d~\cite{Pichler2017} and 3d~\cite{Wan2021} as well as  tree graph states~\cite{PhysRevLett.125.223601}. Such designs involve only a single emitter and relax the multi-emitter requirement. In addition, the photon emission rate is manifestly regular. However, the feedback loop idea by itself relies on a particular photonic qubit encoding (i.e., presence/absence), where single-qubit gates are not readily available. 
Moreover, this type of qubit encoding is very sensitive to dephasing error and may not be a good choice, especially because in these schemes qubits travel through optical fiber delay line. Hence, resource states obtained from this approach cannot be easily manipulated for different tasks without introducing extra overhead. In addition, introducing more than one delay-line~\cite{Wan2021} in this scheme requires multiple ultrafast optical switches for routing which puts an  upper bound on the device overall performance in terms of generation rate.
Needless to say that optical switches lead to larger loss rate in the system.

\begin{figure}
    \centering
    \includegraphics[scale=0.75]{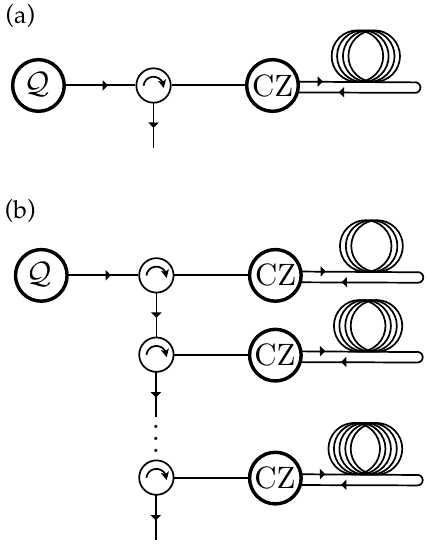}
    \caption{Schematic setup to generate (a) two-dimensional and (b) $N$-dimensional cluster state. $\mathcal{Q}$ and $\cz{}$ denote quantum emitter  and controlled-phase gate (based on a quantum dot array), respectively. Optical fiber delay lines are terminated by a reflective plane; however, we show round-trip trajectories of photons for clarity.}
    \label{fig:cluster-schematic}
\end{figure}

 In this paper, we attempt to resolve the issues with existing prototypes mentioned above and propose a new optical circuit design for deterministic resource state generators.
Our design consists of two components: An active component, a quantum emitter which sends photons to a passive component, comprising an optical circulator attached to a scattering block~\cite{Schrinski2021} with a delay-line feedback loop (see Fig.~\ref{fig:cluster-schematic}). Different graph states can be obtained by changing the resonant pulse sequence applied to QD and possibly additional single-qubit gates. Furthermore, our design is modular, i.e., the passive component can be stacked to realize graphs on higher dimensional lattices. 
We should note that our proposal is not restricted to the QD platform and may be applicable to other quantum emitter technologies. 
Regarding the aforementioned issues, 
our design contains only one emitter; as a result, it does not require any two-qubit gates for emitters and manifestly has a reference clock cycle which is governed by the resonant pulse sequence applied to the emitter. We use time-bin qubit encoding where single-qubit gates are readily available for performing customized tasks on the output resource state.  Additionally, time-bin encoding is generally not sensitive to dephasing error and suitable for long-distance quantum communication~\cite{PhysRevLett.82.2594}.
Last but not least, our setup does not generally require ultra-fast optical switches running at photon generation rate. 
In certain cases, we may have to use an optical switch; however, it only needs to run at a much slower speed than the photon generation rate and hence does not impose a constraint on the overall resource state generation rate.

We provide explicit design protocols for generating cluster states in any dimension, repeater states, and tree graph states. We further discuss an error model for our design where for simplicity we accompany every (effective) gate with a depolarizing channel and calculate the circuit error threshold in the case of fault-tolerant cluster state in three dimensions. We find the circuit error threshold to be $0.53\%$ which shows an improvement compared to the error threshold of $0.39\%$ found in Ref.~\cite{Wan2021}. This roughly means that fault-tolerant error correction can be achieved as along as the gate infidelities are below $0.53\%$.

We also study tolerance against the loss error and {reproduce the known result} that loss rates for directly generating graph states on RHG lattice can be tolerated up to $24.9\%$ (or equivalently $1.24$ dB) corresponding to the bond percolation transition on simple cubic lattice~\cite{Stace-fcts}.
Alternatively, one can imagine generating a 3d cluster state on cubic lattice and then carving out an RHG lattice by measuring a subset of qubits in Z basis. We show the loss threshold in this case to be quite lower around $5.41\%$ (or $0.24$ dB) which is described by the bond percolation transition on cubic lattice with second and fourth nearest neighbor links. 
As we explain, the fault-tolerant regime can be reached despite the loss error in optical fiber delay lines as long as the photon emission rates are greater than $7.5$ or $130$ MHz with or without conversion to the telecomm frequency band.
There are two other places where loss error is dominant, namely, the emitter-waveguide and chip-to-fiber interfaces. 
Chip-to-fiber connection in principle can be improved by engineering a better interface.
At the same time, recent developments in QD technologies have resulted in the emitter radiation efficiency near unity (e.g., $98.4\%$ reported in Ref.~\cite{Arcari2014}) and the path forward looks promising~\cite{Tiurev2020}.

The rest of this paper is organized as follows: 
In Sec.~\ref{sec:preliminary}, we provide some background materials on graph states and the physics of quantum emitters and explain the two  entangling gates available in our setup.  
Next, in Sec.~\ref{sec:design} we present details of our optical circuit design to deterministically generate various graph states. In Sec.~\ref{sec:error}, we focus on fault-tolerant properties of 3d cluster states and show that our optical circuit can generate a robust state in the presence of photon loss and gate errors. Finally, we close in Sec.~\ref{sec:conclusions} with concluding remarks and future directions. Some of the details and derivations are presented in three appendices.


\section{Preliminary remarks}
\label{sec:preliminary}

In this section, we review some basics of the graph states and also discuss the building blocks of our optical circuits, namely, quantum dots coupled to a waveguide which we use to generate entangled photons and implement deterministic gates.

\subsection{Graph state representation}

\begin{figure}
    \centering
    \includegraphics[scale=0.9]{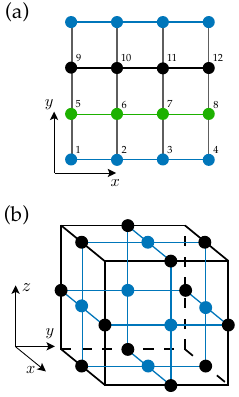}
    \caption{(a) 2-dim cluster state. Qubits are numbered and color coded according to the circuit in Fig.~\ref{fig:cluster-circuit}.
 (b) RHG lattice for fault-tolerant quantum computing. In panel (b), black edges are to used to show the 3D structure, while blue edges represent edges in the graph state representation.}
    \label{fig:clusterstate}
\end{figure}

In this part, we review some definitions and introduce our notations.
Consider an undirected graph $G$ with $N$ vertices representing $N$ qubits.
A graph state~(see Ref.~\cite{Hein-review} for a detailed review) associated with $G$ is then defined as a quantum state of $N$ qubits and given by
\begin{align}
    \ket{\Psi_G} =\prod_{(i,j)\in G} \cz{i,j} \ket{+}^{\otimes N},
\end{align}
where subscripts are qubit labels $i,j=1,\cdots,N$, and $\cz{i,j}$ is a controlled-phase gate between qubits $i$ and $j$ for every edge $(i,j)$ on graph $G$. 
We should note that the ordering of $\cz{}$ gates in the above expression does not matter as they commute with each other. Throughout this paper, we use $\{ X_i,Y_i,Z_i\}$ to denote the Pauli operators acting on qubit $i$. Also, $\ket{+}$ refers to the eigenstate of $X$ Pauli operator, i.e., $X\ket{\pm}=\pm\ket{\pm}$, which can be obtained by a Hadamard gate $\ket{+}=H\ket{0}$.

An important property of graph states is that they can be characterized as stabilizer states, i.e., a common eigenvector of a stabilizer group generated by $N$ commuting set of Pauli operators (aka stabilizer generators) $\mathcal{S} = \{ P_1, \cdots, P_N
\}$. The $i$-th stabilizer generator associated with the $i$-th vertex on $G$ is defined by
\begin{align}
    P_i = X_i \bigotimes_{(i,j)\in G} Z_j.
\end{align}
In other words, the $i$-th stabilizer is a product of an $X$ Pauli operator on $i$-th vertex and $Z$ Pauli operators of adjacent (in the sense of graph) vertices. We should note that graph state is a stabilizer state (as opposed to a stabilizer code) since there are $N$ stabilizers which determine a unique state for $N$ qubits.

\begin{figure}
    \centering
    \includegraphics[scale=0.5]{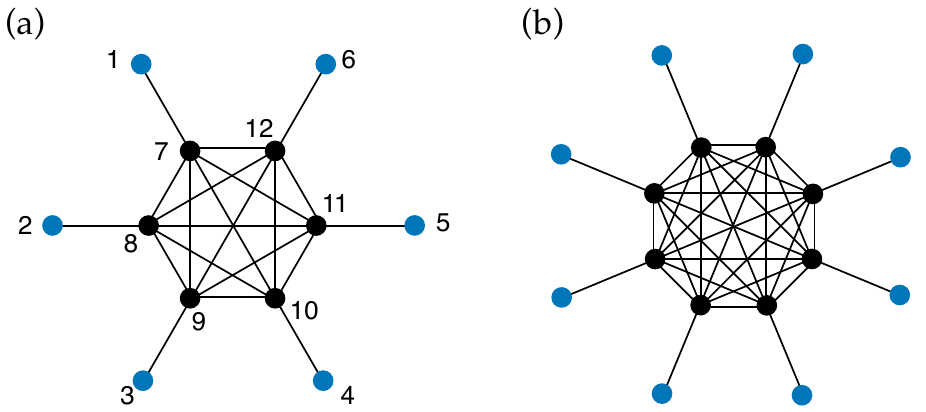}
    \caption{Graph representation of photonic repeater states of (a) twelve and (b) sixteen qubits.  Numbers indicate the temporal position of the photon in the sequence.}
    \label{fig:all-to-all}
\end{figure}

\begin{figure}
    \centering
    \includegraphics[scale=0.5]{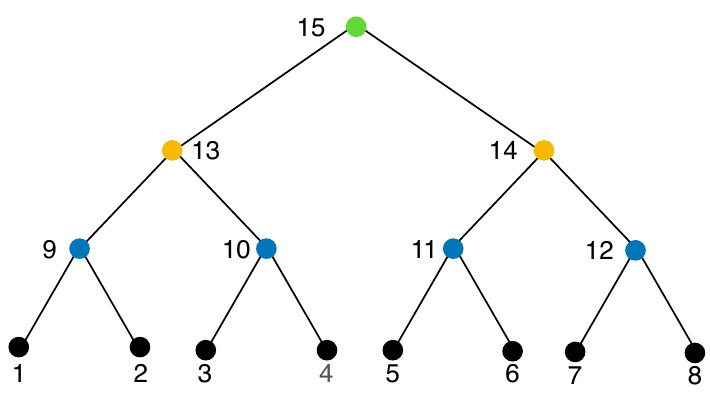}
    \caption{Tree graph state of fifteen qubits. Numbers indicate the temporal position of the photon in the sequence.}
    \label{fig:tree}
\end{figure}

There are several advantages to use graph states; to name two, let us mention that they are equipped with a graphical representation which gives some intuition to the multipartite entanglement nature of the system, and second the stabilizer formalism provides an efficient classical description of the graph state (despite being a non-trivial many-body state) which makes the simulation of error processes (and possible error correction schemes) tractable on a classical computer.
Figures~\ref{fig:clusterstate}-\ref{fig:tree} show some commonly used graph states with various applications from quantum computing to quantum networking and communications. Graph states on a hypercubic lattice in $N$-dimensions are generally referred to as $N$-dim cluster states (see Fig.~\ref{fig:clusterstate}). A 1d (linear) cluster state is a simplest state which can be used as a teleportation channel or for entanglement distribution in a quantum network. A 2d cluster state shown in Fig.~\ref{fig:clusterstate}(a) is a resource state for universal quantum computation in a measurement-based computation scheme~\cite{Raussendorf-2d,PhysRevLett.86.5188}. Most importantly, a 3d cluster state or its variant in the form of a Raussendorf-Harrington-Goyal (RHG) lattice~\cite{Raussendorf-ftcs,PhysRevA.71.062313} (Fig.~\ref{fig:clusterstate}(b)) can be used as a platform for universal fault-tolerant quantum computation.

Other useful graph states with applications in quantum communication include all-to-all  graph (repeater) states such as the ones shown in Fig.~\ref{fig:all-to-all} 
 for all photonic quantum repeaters~\cite{Azuma2015} and tree graph states as a quantum error-correcting code to realize a one-way quantum communication protocol~\cite{Borregaard2020}.

Before, we delve into the details of our optical circuit designs, let us briefly go over the circuit building blocks, i.e., quantum emitters and controlled-phase gates in the next part.

\subsection{Quantum emitters for deterministic gates}

\label{sec:deterministic gates}

As mentioned, quantum emitters such as a quantum dot coupled to a waveguide can be used both as a deterministic source of entangled photons and a nonlinear medium for two-photon gates. In this section, we review these facts in some detail which then will be used in the following sections as principles for the circuit design.
As we see, the photon emission process is effectively a controlled-not gate where the control qubit is the emitter and the target qubit is a photonic qubit. 

\begin{figure}
\centering
\includegraphics[scale=0.9]{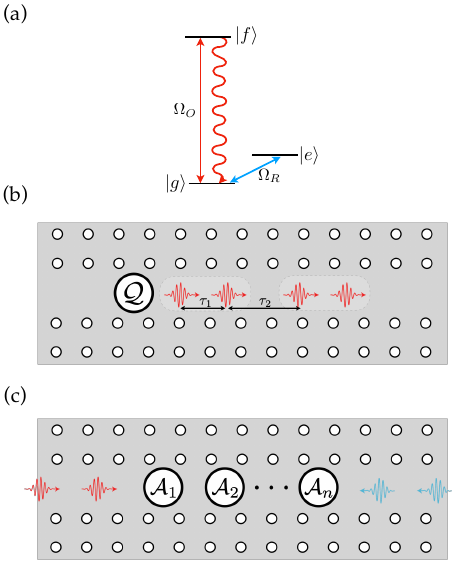}
\caption{(a) Energy levels of the quantum dot (emitter) where we use $\{\ket{e},\ket{g}\}$ as the computational basis for the emitter. (b) Quantum dot coupled to photonic waveguide as a photon source. We use time-bin encoding characterized by intra-qubit time-delay $\tau_1$ and inter-qubit time difference of $\tau_2$. (c) Array of quantum dots to mediate a controlled-phase gate between two time-bin photonic qubits incident from left and right.
\label{fig:emitter}}
\end{figure}

Throughout this paper, we use an emitter with a $\Lambda$-type energy level structure as shown in Fig.~\ref{fig:emitter}(a),  where we use the transition $\ket{g}\leftrightarrow \ket{f}$ to send and absorb photons denoted by $a$ and $a^\dag$. 
The other state is denoted by $\ket{e}$ and the transition is $\ket{e} \leftrightarrow \ket{f}$ is not active and undesired. The logical states of the emitter are 
\begin{align}
\ket{1}_q \coloneqq \ket{g}, \qquad    \ket{0}_q \coloneqq\ket{e}.
\end{align}
Such energy levels can be realized in a QD subject to a few Tesla magnetic field~(see e.g.~\cite{Appel2021}), where
$\ket{g}$ and $\ket{e}$ correspond to electron or hole spin degree of freedom, and $\ket{f}$ describes a state with an additional exciton. 

We apply two types of resonant pulses: One that controls the logical state of the emitter $\ket{g}\leftrightarrow \ket{f}$ with Rabi frequency $\Omega_O$, and another one that drives the transition  $\ket{g}\leftrightarrow \ket{e}$ with Rabi frequency  $\Omega_R$. We should note that in the case of QDs the latter transition (between electron or hole spin up and down) is 
driven by a
detuned Raman laser and is 
not an optical dipole transition~\cite{Appel2021}; nevertheless, this process results in a transition with an effective ground state coupling $\Omega_R$.

For the photonic qubits, 
we use a time-bin encoding of quantum information based on the photon arrival (generation) time.
The logical states of a photonic qubit are defined by 
\begin{align}
\label{eq:time-bin}
\ket{1}_p\coloneqq a_e^\dag \vac, \qquad \ket{0}_p\coloneqq a^\dag_l \vac,    
\end{align}
where the subscript of the photon creation operator $e\ (l)$ denotes the early (late) generation time, respectively. Here, $\vac$ denotes the vacuum state of the photonic mode in the waveguide. 
Two advantages of using this encoding scheme is that single-qubit gates can be implemented and dephasing error is nearly supressed. A potential challenge with this encoding would be implementing a controlled-phase gate between the emitter and photonic qubits, where a simple scattering off the emitter~\cite{Lodahl-chiral-2017,Pichler2017} is not effective since we get a $\pi$-phase shift regardless of the logical state of the photonic qubit. However, as we explain shortly, we can implement a controlled-phase gate between two time-bin encoded photonic qubits by using an array of QDs following a recent proposal by Schrinski \emph{et~al.}~\cite{Schrinski2021}.
In what follows, we explain the two basic gates we use to generate entangled states: A controlled-not gate during the emission process and a controlled-phase gate through a scattering process.

\emph{Emission process:} 
We initialize the emitter state in a Hadamard state $\ket{+}_q = (\ket{0}_q+\ket{1}_q)/\sqrt{2}$ by sending a resonant $\pi/2$-pulse between $\ket{0}\leftrightarrow \ket{1}$ at $\Omega_R$.
This corresponds to a Hadamard gate on the emitter, which we denote by $H_q$.
Then, we send a resonant $\pi$-pulse of $\Omega_O$ which after possible spontaneous emission gives the following state
\begin{align}
\label{eq:bell presence}
 & \frac{1}{\sqrt{2}} ( \ket{1}_q \otimes a_e^\dag \vac + \ket{0}_q \otimes \vac) \nonumber.
\end{align}
Note that the photon emitted at this instance is the early one.
Next, we apply a $\pi$-pulse of $\Omega_R$ followed by another  $\pi$-pulse of $\Omega_O$ where the state after the emission is found to be
\begin{align}
 & \frac{1}{\sqrt{2}} ( \ket{0}_q \otimes a_e^\dag \vac + \ket{1}_q\otimes  a_l^\dag \vac) \nonumber \\
&= \frac{1}{\sqrt{2}} ( \ket{0}_q \otimes \ket{1}_p + \ket{1}_q \otimes \ket{0}_p),
\end{align}
After another $\pi$-pulse of $\Omega_R$ we obtain the desired state
\begin{align}
\label{eq:bell el}
 \frac{1}{\sqrt{2}} ( \ket{1}_q\otimes \ket{1}_p + \ket{0}_q \otimes \ket{0}_p),
\end{align}
which is a Bell pair between the emitter and the photonic qubit.
Therefore, 
the emission process effectively acts as a $\cnot{q,k}$ gate where the control qubit is the emitter. Combined with the emitter initialization, this process corresponds to the following quantum circuit
\begin{align}
\label{eq:emission}
\makebox[0pt][c]{\raisebox{-2.5ex}{\includegraphics[scale=1]{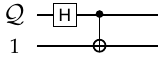}}}
\end{align}
As we discuss in the next section, by applying resonant pulses regularly we generate an array of time-ordered photonic qubits in a desired graph state. 
In particular, we use the emitter (possibly equipped with single-qubit gates) to generate linear cluster states, star graph states, and single qubits in $\ket{\pm}$ states.
The basic timing requirement is 
$\gamma \tau_i\gg 1 $ where $\tau_i^{-1}$ is the photon generation rate (or $\tau_i$ with $i=1,2$ is the time difference between two consecutive photons, see Fig.~\ref{fig:emitter}(b)), and $\gamma$ is the emitter decay rate. Hence, a QD with $\gamma\sim 1$-$100$ GHz~\cite{Appel2021,Tiurev2020} can accommodate photon emission rates of $\tau\sim 0.1$-$10$ nsec.

\begin{figure*}
    \centering
    \includegraphics[scale=0.75]{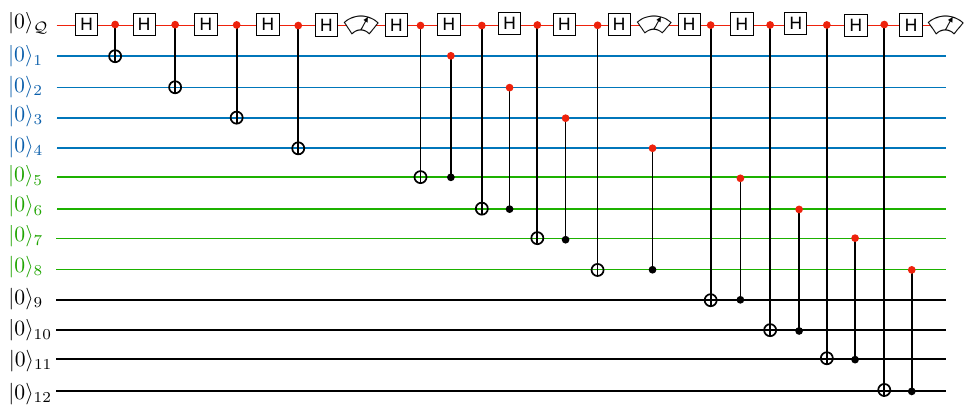}
    \caption{Effective quantum circuit to generate 2D cluster state (Fig.~\ref{fig:clusterstate}(a)) using the setup shown in Fig.~\ref{fig:cluster-schematic}, where qubit numbers indicate the temporal position of the photonic qubit in the sequence.}
    \label{fig:cluster-circuit}
\end{figure*}

\emph{Scattering process:} To build arbitrary graph states, we need another two-qubit gate besides the nonlinearity during the emission process. This could be achieved by using the fact that a photon scattered by a (resonant) QD accumulates a $\pi$-phase shift~\cite{Shen:05,PhysRevA.76.062709,PhysRevLett.114.173603}, which effectively acts as a $\cz{}$ gate between the QD and the photonic qubit (in a presence/absence encoding)~\cite{Pichler2017}. 
A possible circuit design for arbitrary graph states would then have delay-line feedback loops into a single quantum emitter~\cite{Pichler2017,Wan2021}. Despite the simplicity and minimal requirement for number of components, these designs need high speed optical switches as fast as $\tau^{-1}$ (photon generation rate) for photon routing. Furthermore, it is not clear how to implement an arbitrary single-qubit gate for universal quantum computation in a presence/absence encoding.
As mentioned, we would like to use the time-bin encoding, as a result of which the above emitter-photonic-qubit $\cz{}$ gate is not available (simply because in either logical state of photonic qubit there will be a $\pi$-phase shift).
Hence, to induce a $\pi$-phase shift, we use the nonlinearity induced by an array of QDs as shown in Fig.~\ref{fig:emitter}(c) following Ref.~\cite{Schrinski2021} {(We note that these theoretical designs are yet to be realized in the lab)}. The basic idea is based on the absence of scattering $\pi$-phase shift when two photons simultaneously pass through a QD. If the time-bin photonic qubits are scheduled to arrive such that the late component of the incoming photon from left overlaps with the early component of the incoming photon from right, then we will get the following (relative) phase-shifts
\begin{align}
    \ket{e_L,e_R} &\to  \ket{e_L,e_R}, \nonumber \\
    \ket{l_L,e_R} &\to  -\ket{l_L,e_R}, \nonumber \\
    \ket{e_L,l_R} &\to  \ket{e_L,l_R}, \nonumber \\
    \ket{l_L,l_R} &\to  \ket{l_L,l_R}.
\end{align}
Here, $\ket{t_s}\coloneqq a^\dag_{ts}\vac$ in which $t = e\ (l)$ refers to early (late) time-bin in photonic qubits and $s=L\ (R)$ denotes the incoming qubit from left (right) (c.f.~Fig.~\ref{fig:emitter}(c)).
From the computational basis defined in Eq.~(\ref{eq:time-bin}), it is easy to see that the above transformation corresponds to a combined gate $X_L \cz{L,R} X_L$.

In the next section, we introduce our designs by combining a quantum emitter and $\cz{}$ gates along with delay lines to generate various graph states.

\begin{figure*}
    \centering
    \includegraphics[scale=0.75]{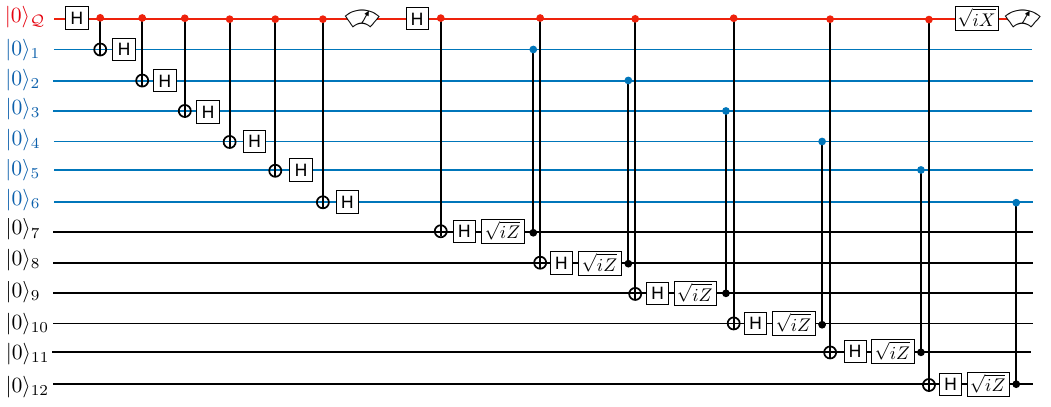}
    \caption{Effective quantum circuit to generate 12-qubit repeater state in Fig.~\ref{fig:all-to-all}(a) using the setup in Fig.~\ref{fig:repeater-schematic}. Here, numbers indicate the temporal position of the photon in the sequence.}
    \label{fig:repeater-circuit}
\end{figure*}

\begin{figure}
    \centering
    \includegraphics[scale=0.75]{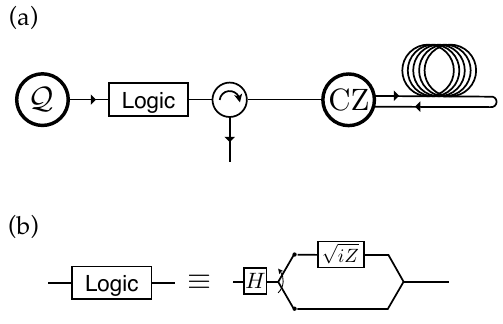}
    \caption{Schematic setup to generate a repeater graph state (Fig.~\ref{fig:all-to-all}). $\mathcal{Q}$ and $\cz{}$ denote quantum emitter  and controlled-phase gate (based on a quantum dot array), respectively. Panel (b) shows the contents of logic gate block in (a). We note that to generate a repeater state of $2N$ qubits, the switch remains at each position for $N\tau$.}
    \label{fig:repeater-schematic}
\end{figure}

\section{Optical Circuit Designs for Graph States}
\label{sec:design}

Our typical optical circuit design to generate time-ordered graph states is shown in Fig.~\ref{fig:cluster-schematic}(a). 
It consists of an active component and a passive component which may be composed of a few passive modules. The active component is a quantum emitter, denoted by $\mathcal{Q}$, coupled to a waveguide as a source of either linear clusters or individual photonic qubits. Each passive module consists of an optical circulator, a $\cz{}$ (scattering) gate, and an optical fiber delay-line that is in turn terminated by a reflector (or mirror).
The entanglement is built within both passive and active regions of the circuit.
As we explain later in this section, passive modules can be stacked to generate more complex graph states (see Fig.~\ref{fig:cluster-schematic}(b)). 
Depending on the type of graph state we wish to generate, we apply a series of $\Omega_O$ and $\Omega_R$ pulses to the $\mathcal{Q}$ emitter and may need to have additional single-qubit gates as shown in Fig.~\ref{fig:repeater-schematic}.

\begin{figure*}
    \centering
    \includegraphics[scale=0.75]{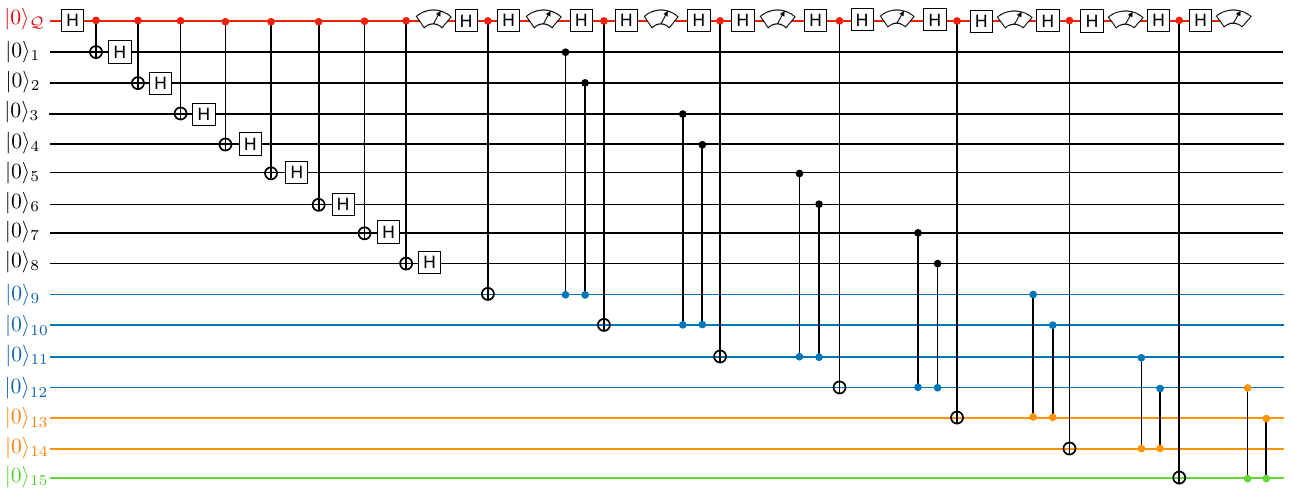}
    \caption{Effective quantum circuit to generate 15-qubit binary tree state in Fig.~\ref{fig:tree} using the setup in Fig.~\ref{fig:cluster-schematic}(a).}
    \label{fig:tree-circuit}
\end{figure*}

In what follows we shall use two basic subgraphs, namely, the star graph and linear cluster, generated by a quantum emitter to form more complex graphs.
As we recall in Sec.~\ref{sec:deterministic gates}, a five-step sequence of $\Omega_O$ and $\Omega_R$ resonant pulses lead to a Bell-pair between the emitter and a photonic qubit: (i) $\frac{\pi}{2}$-pulse of $\Omega_R$, (ii)-(v) four alternating $\pi$-pulses of $\Omega_O$ and $\Omega_R$.
{Repeating the steps (ii)---(v)} $N$-times
  lead to a GHZ state of $N$ photonic qubits and the emitter. Upon applying hadamard to photonic qubits, we obtain a star graph,
\begin{align}
    \label{eq:star-graph}
    \makebox[0pt][c]{\raisebox{-2.5ex}{\includegraphics[scale=0.7]{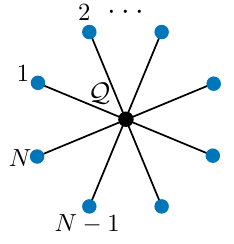}}}
\end{align}
where the central vertex on the graph is the emitter.
The effective circuit is depicted by:
\begin{align}
    \label{eq:ghz-circuit}
    \makebox[0pt][c]{\raisebox{-2.5ex}{\includegraphics[scale=0.8]{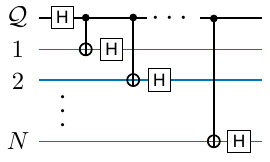}}}
\end{align}
Repeating all five steps lead to a linear cluster state such that in the graph representation the emitter is adjacent to the last emitted photonic qubit as in
\begin{align}
    \makebox[0pt][c]{\raisebox{-2.5ex}{\includegraphics[scale=0.85]{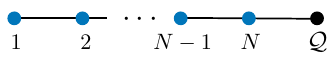}}}
\end{align}
It is worth noting that we can combine the last $\pi$-pulse of $\Omega_R$ in a given step with the first $\frac{\pi}{2}$-pulse of $\Omega_R$ in the next step to make it a four-step pulse sequence: Three $\pi$-pluses of $\Omega_O$, $\Omega_R$, and $\Omega_O$, respectively, along with a $\frac{\pi}{2}$ $\Omega_R$ pulse. 
The corresponding quantum circuit can be shown as
\begin{align}
    \label{eq:linearcluster-circuit}
    \makebox[0pt][c]{\raisebox{-2.5ex}{\includegraphics[scale=0.85]{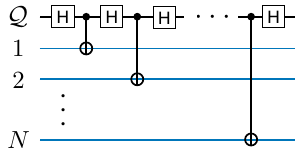}}}
\end{align}
The above two protocols are modified versions of the original Lindner-Rudolph~\cite{Lindner-Rudolph2009} machine-gun proposal to account for qubit time-bin encoding. We denote the time difference between early and late bins within a photonic qubit by $\tau_1$ and the time difference between the late and early bins of two consecutive qubits by $\tau_2$ (see Fig.~\ref{fig:emitter}(b)). This simply implies that the time difference between the early bins of two consecutive qubits is $\tau = \tau_1+\tau_2$. As we see, we can obtain graph states on different lattices by tuning the $\tau_1/\tau_2$ ratio.

The other subgraph we may need is an array of unentangled qubits prepared in hadamard basis $\ket{\pm}$. We propose three ways to do that.
 One way to achieve this is to interrupt the linear cluster generation by measuring the emitter in the Hadamard basis (which puts the photonic qubit in $\ket{\pm}$ depending on the measurement outcome). This corresponds to the following circuit:
\begin{align}
\label{eq:photon-hadamard}
\makebox[0pt][c]{\raisebox{-2.5ex}{\includegraphics[scale=0.9]{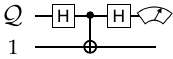}}}
\end{align}
Another way is to prepare a star graph (\ref{eq:star-graph}) and measure the emitter afterwards.
As a third method, we can put the emitter in the logical state $\ket{1}$ and run it as a two-level system. Then, applying a sequence of $\pi$-pulses of $\Omega_O$ yields a series of single photons.
By placing a Mach-Zehnder interferometer (MZI) with a phase-shifter and a delay line (of $\tau_1$) 
\begin{align}
    \label{eq:MZI}
    \makebox[0pt][c]{\raisebox{-2.5ex}{\includegraphics[scale=0.5]{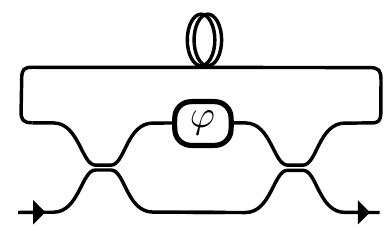}}}
\end{align}
we can then generate a sequence of time-bin qubits in $\ket{+}$ as follows: An early photon passes through MZI while we set $\varphi=\frac{\pi}{2}$ so that half of the incoming signal goes to the delay line and the other half leaves. Upon approaching $t=\tau_1$ we set $\varphi=\pi$ so that the output of the delay line comes out of the MZI.
We should note that these three methods have their own pros and cons. For example, the emitter measurement and reinitialization could be slow ($\sim 0.1$-$1\ \mu$sec)~\cite{Appel2021}, which is used for every qubit in method one. In contrast, the latter two methods only require emitter resetting once. However, they come at the cost of single qubit gates such as Hadamard in the second method (\ref{eq:ghz-circuit}) and MZI with a delay line in the third method (\ref{eq:MZI})
which would introduce further loss into the system.

\subsection{N-dimensional cluster states}

\label{sec:cluster states}

Our algorithm to make higher dimensional cluster states is incremental in the sense that we start with linear cluster states and stitch them together via $\cz{}$ scattering gate to get a 2d cluster state (Fig.~\ref{fig:clusterstate}(a)) at the output. This step involves keeping $n$-th linear cluster in the delay line while $(n+1)$-th line is being prepared by the emitter as shown in Fig.~\ref{fig:cluster-schematic}(a). As a result, if we want to produce a $L_x \times L_y$ 2d cluster state (qubit time-orderings and their coordinates are illustrated in Fig.~\ref{fig:clusterstate}(a)), we need a delay line of $L_x (\tau_1+\tau_2)+\tau_1$ (such that $\tau_1<\tau_2$) and continue the process $L_y$ times. The effective quantum circuit is shown in Fig.~\ref{fig:cluster-circuit}. We should note that here we measure the emitter after the completion of each line so that we get a 2d cluster state with open boundaries. Setting $\tau_1=\tau_2$, we obtain a 2d graph state on triangular lattice.
To generate a 3d cluster state, we add another passive module at the output of this circuit which is used to stitch together layers of 2d cluster states. To make a $L_x\times L_y\times L_z$ lattice, we need to have a $L_x L_y (\tau_1+\tau_2)+\tau_1$ delay line in the second module. Repeating this process over $N$-modules lead to an $N$-dimensional graph state on a hypercubic lattice.

An important example is the RHG state~\cite{Raussendorf-ftcs,Raussendorf_2007} shown in Fig.~\ref{fig:clusterstate}(b) which can be obtained by measuring (in $Z$-basis) a subset of photons on 3d cubic lattice, which form a body-centered lattice in a doubled unit cell.
Alternatively, we can make an RHG state by modifying the resonant pulse sequence in every other line and use one of the three methods discussed near the end of the previous part (see the discussion around Eq.~(\ref{eq:photon-hadamard})) so that we get a sequence of qubits in $\ket{+}$. 
This gives the downward output of the first optical circulator to be the desired state on $xy$ plane of RHG state, which is a cluster state version of surface code with half of stabilizers.
As we will see in the next section, preparing a cubic lattice and measuring qubits has a lower loss threshold compared to the direct construction (i.e., the three aforementioned methods); however, the former method does not involve any switches or resetting the emitter, and hence, might have a better performance in terms of state generation rate since resetting emitters (in particular QDs) is potentially a slow process.

Let us now compare our protocol with the previous studies~\cite{Pichler2017,Wan2021}. First, both single-rail encodings in previous work are based on absence/presence encoding where applying a direct arbitrary single-qubit gates is not possible. This is particularly important because the most basic requirement is measurement in $X$ basis (e.g., for error correction) which is not immediately available in such encoding. In contrast, our protocol is based on time-bin encoding which can be easily converted to dual-rail encoding where linear optics can be used for single-qubit gates. The dual-rail encoding in Ref.~\cite{Wan2021} involves distinguishable photons and may be used for applying arbitrary single-qubit gates, but at the cost of extra overhead such as frequency converters. 
Second, adding two delay-lines~\cite{Wan2021} requires ultrafast switching at the photon emission rate, while there is no need for ultrafast switches or only one switch in our scheme.
Third, the RHG generation protocol in Ref.~\cite{Wan2021} requires emitter resetting quite frequently. As mentioned above, this is potentially a very slow process for QDs and not experimentally desirable. In our case, we have two alternative ways of doing that by using a star graph or running the QD as a two-level system (albeit at the cost of introducing actively reconfigured MZI switches), whereas there is no such flexibility in Ref.~\cite{Wan2021}.


\subsection{Repeater graph states}

One of the challenges to form an entanglement distribution across a quantum network is amplifying a quantum signal which is prohibited by the no-cloning theorem. A possible solution is to realize long-range Bell-pairs by preparing local Bell-pairs and sequentially send one (photonic) qubit of each repeater station to the next station where a fusion measurement gate is performed. The issue with this approach is that fusion gates are probabilistic and over a long distance (and many repeater stations) the success probability of generating a long-range Bell-pair decays (exponentially) rather quickly. To increase the success probability of each fusion gate, all-photonic repeater states were developed~\cite{Azuma2015}, which effectively act as a multiplexed fusion gate: Instead of sending one qubit of Bell-pair, multiple qubits (left/right half of repeater states in Fig.~\ref{fig:all-to-all}) are sent and multiple fusion gates are performed on the outer qubits (labeled 1 to 6 in Fig.~\ref{fig:all-to-all}(a)). Depending on the fusion outcomes, inner qubits are measured to complete a Bell link between two consecutive repeater station.
Since the pioneering work~\cite{Azuma2015}, there have been several studies on trade-off and resource requirements to realize these states deterministically~\cite{Economou2017,Russo_2019,Hilaire2021resource,Li2021,PhysRevLett.125.223601}
 and perform various quantum communication tasks based on the repeater states~\cite{PhysRevA.95.012304,Azuma2015-qkd,Zwerger2016}.

Here, as a byproduct of our design, we show how to determinstically generate repeater states in our setup. We should note that compared to previous works~\cite{Economou2017,Russo_2019,Hilaire2021resource,Li2021} which use actively controlled multiple QDs (or solid-state emitters) and require two-qubit gates on them, our setup has only one active component. Figure~\ref{fig:repeater-schematic} shows a schematic optical circuit which only differs from the 2d-cluster state generator of Fig.~\ref{fig:cluster-schematic}(a) by a logic gate block.

We now explain how to generate repeater state of $2N$ qubits (see Fig.~\ref{fig:all-to-all}(a) and (b) for $N=6$ and $8$, respectively).
As mentioned, the multi-qubit output state in our system is time-ordered. A natural ordering for repeater states is to enumerate outer qubits and then inner qubits (e.g., Fig.~\ref{fig:all-to-all}(a)). Given this ordering, we need to define a sequence of resonant pulses on the quantum emitter and a proper delay line so that the outer qubits in $\ket{\pm}$ states and collide them with an all-to-all connected graph of inner qubits.
The equivalent quantum circuit is shown in Fig.~\ref{fig:repeater-circuit}.
To obtain the $N$ outer qubits, we have three options as we discussed earlier in this section. In the quantum circuit depicted in Fig.~\ref{fig:repeater-circuit}, we use the star graph (\ref{eq:star-graph}) and then measure the emitter. 
The generated sequence of $N$ outer qubits goes through the identity path in the logic block of Fig.~\ref{fig:repeater-schematic}(b) and is held in the delay line while the inner part is prepared. Hence, the duration of the delay line needs to be $N(\tau_1+\tau_2)+\tau_1$ (provided that $\tau_1<\tau_2$), and 
the optical switch rate is $N(\tau_1+\tau_2)$ since we only need to have an additional single qubit gate for the inner qubits.
For the inner layer, we prepare another star graph and apply a local complementation~\cite{PhysRevA.69.022316,PhysRevA.69.062311} with respect to the emitter vertex and measure the emitter in $Z$ basis subsequently. The local complementation is achieved by $\sqrt{iZ}$ and $\sqrt{iX}$ gates on qubits and the emitter, respectively. It is worth noting that the single-qubit gates commute with each other and only need to be performed before the $\cz{}$ gate.
Next, the all-to-all graph is stitched to the outer qubits in the delay line 
by the $\cz{}$ scattering gate. We also note that since our design is modular we can stack another passive layer at the output of this circuit (similar to \ref{fig:cluster-schematic}(b)) and effectively add multiple levels of tree branches to the outer qubits (see next part below for further details).

\subsection{Tree graph states}

One of the main challenges in one-way quantum communication with discrete variable (DV) encoding is photon loss. A possible resolution within the DV formalism is to use quantum error correcting codes~\cite{PhysRevLett.97.120501}. This way, quantum information is encoded in several qubits in the form of a stabilizer state and sent to a repeater station where the state is reconstructed (even though a portion of qubits are lost) based on the remaining qubits received at that point. As it was shown in Ref.~\cite{PhysRevLett.97.120501}, tree graphs could be a good candidate stabilizer state for this purpose.
Another application of tree states is a tree-type channel between two nodes on a graph as a loss tolerant teleportation channel~\cite{Morley_Short_2019}.

Let us now explain how to generate a binary tree graph of $N$ layers consisting of $2^N-1$ qubits.
We consider a natural ordering of vertices from bottom to top as shown in Fig.~\ref{fig:tree} for $N=4$. The resonant pulse sequence we develop is applied to an optical circuit similar to Fig.~\ref{fig:cluster-schematic} and realizes the quantum circuit shown in Fig.~\ref{fig:tree-circuit}. 
All the qubits here are generated by using the single-qubit protocol in Eq.~(\ref{eq:photon-hadamard}). Similar to the repeater states, the lowest layer can alternatively be prepared by generating a star graph state and measuring the emitter. The delay line and pulse sequence is designed such that upper level qubits collide with two consecutive lower level qubits while passing through $\cz{}$ and realize a binary branch structure. For instance, late component of qubit $1$ and early component of qubit $9$ and early component of qubit $2$ and late component of qubit $9$ meet at $\cz{}$ scattering center. {Therefore, we need to use a varying time-bin encoding in each layer (according to a pattern which we discuss below) and include a superscript $(n)$ to denote the layer number in the tree structure from $n=0$ (lowest) to $n=N-1$ (highest).} By a direct time-analysis we find that
\begin{align}
    \label{eq:tree-diff-eq}
    \tau_1^{(n+1)} &= \tau_2^{(n)}, \nonumber \\
    \tau_2^{(n+1)} &= 2\tau_1^{(n)}+ \tau_2^{(n)},
\end{align}
{We assume $\tau_1^{(0)}=\tau_2^{(0)}$ which implies the delay line duration to be $\Delta = (2^N-1)\tau_1^{(0)}$.}
Equation~(\ref{eq:tree-diff-eq}) leads to the following temporal bins for the $n$-th layer
\begin{align}
    \tau_1^{(n)} &= \frac{\tau_1^{(0)}}{3}\left(2^{n+1}+(-1)^n\right), \nonumber \\
    \tau_2^{(n)} &= \frac{\tau_1^{(0)}}{3}\left(2^{n+2}-(-1)^n\right).
\end{align}
The emission time of the first qubit in the $n$-th layer satisfies the following relation 
\begin{align}
    t_{\text{init}}^{(n+1)} = \Delta + t_{\text{init}}^{(n)} + \tau_1^{(n)},
\end{align}
and the solution is given by
\begin{align}
    t_{\text{init}}^{(n)} &= n\Delta + \sum_{j=0}^{n-1} \tau_1^{(j)} \\
    &= n\Delta + \frac{\tau_1^{(0)}}{6}\left( 4(2^{n}-1)+1-(-1)^n \right),
\end{align}
where we set the time origin at the moment when the early component of the qubit $1$ is emitted, i.e., $t_{\text{init}}^{(0)}=0$.

In closing, let us note that all three types of graph states discussed in this section can be obtained fully by a sequence of qubits in $\ket{+}$ and a $\cz{}$-delay line block.
Hence, we may use a regular QD (two-level system) as a single-photon emitter and a linear optical circuit of Eq.~(\ref{eq:MZI}) to generate qubits in $\ket{+}$ states using time-bin encoding.

\begin{figure}
    \centering
    \includegraphics[scale=0.55]{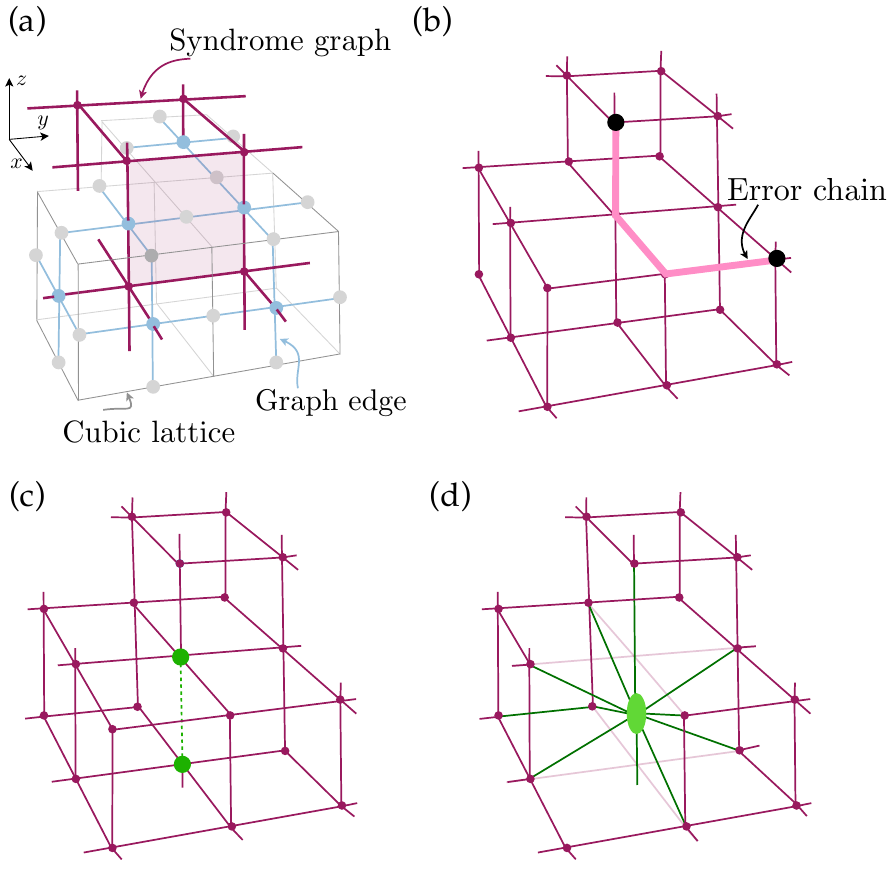}
    \caption{(a) Syndrome graph shown in violet, where vertices and edges are associated with parity checks (cells of RHG cubic lattice) and qubits (faces), respectively, superimposed on RHG lattice shown in gray. Bit flip errors are shown in (b) and a loss error in (c) and (d).
    (b) An error chain (pink) with two failed parity checks at its endpoints (black vertices). As explained in the text, stabilizer generators need to be modified to account for a qubit loss. (c) A lost qubit is shown by the dashed green edge where the two adjacent parity checks are impacted. (d) By combining the two impacted stabilizers, a supercheck operator (green ellipse) is constructed and the syndrome graph is no longer a regular (simple cubic) lattice. }
    \label{fig:loss-syndrome}
\end{figure} 
\section{Error analysis of fault-tolerant state}
\label{sec:error}

In this section, we focus on a graph state on an RHG lattice which is a candidate state for fault-tolerant quantum computing and show that our protocol can realize a fault-tolerant regime. 
Before we delve into details of error models, let us briefly review how error correction works in an RHG cluster state. Here, we only consider the cluster state as a quantum memory. 

As shown in Fig.~\ref{fig:clusterstate}(b), the graph representation defines face stabilizers as
\begin{align}
    S_f = X_f \bigotimes_{e\in \partial f} Z_e,
\end{align}
where $f$ denotes a (primal) cubic face (black squares in Fig.~\ref{fig:clusterstate}(b)) and $\partial f$ denotes the boundary edges of face $f$. A subgroup of the stabilizers can be constructed in terms of product of the face stabilizers $S_f$ corresponding to the faces of a given cell $c$ which takes the form of
\begin{align}
\label{eq:cube}
    S_c = \bigotimes_{f\in \partial c} S_f= \bigotimes_{f\in \partial c} X_f,
\end{align}
where the edge Pauli operators cancel out.
Here, $\partial c$ denotes the boundary faces of cell $c$. Another subgroup of stabilizers can be similarly defined in terms product of the stabilizers associated with edges connected to a vertex on the primal (or cells on the dual) lattice. The two set of stabilizers can be used towards a topological quantum error correction as explained below.

\subsection{Pauli error threshold}

In the absence of loss, error correction process on primal lattice (which in this case is simple cubic) is realized by measuring qubits on faces in $X$ basis and calculating the parity checks associated with cell stabilizers (\ref{eq:cube})~\cite{Raussendorf-ftcs}. It is more convenient to consider the dual lattice, where cell stabilizers are defined on vertices and qubits are placed on the edges. Such a lattice is called a syndrome graph (Fig.~\ref{fig:loss-syndrome}(a)). The erroneous qubits on the syndrome graph form an error chain connecting failed parity checks (a.k.a.~syndrome defects), see Fig.~\ref{fig:loss-syndrome}(b). The goal of error correction is to find recovery (or correction) chains which match the syndrome defects as their endpoints. By definition, the error and recovery chains form closed loops. Logical errors occur when these loop operators do not commute with logical operators, otherwise, the error correction succeeds.

In practice, finding a suitable correction chain is computationally difficult. Following the literature~\cite{WANG200331,Raussendorf-ftcs}, we use Edmonds' minimum-weight perfect matching (MWPM) algorithm~\cite{edmonds_1965} which is known as a heuristic but efficient method for this purpose. The MWPM algorithm, in short, pairs up syndrome defects which are closest to each other and minimize the sum of  weights of the correction chains, i.e., minimum number of qubits connecting failed checks (which, on a regular lattice with equal error rate on each link, simply equals the taxicab distance, c.f.~Fig.~\ref{fig:loss-syndrome}(b)).
As we explain below, the effective error rate of physical qubits in our setup is not uniform since the state generation process is anisotropic. Because of that, we use Dijkstra’s algorithm for finding shortest paths on a weighted graph. 
In our numerical simulations, we wish to probe the bulk properties; hence, we consider periodic boundary conditions in all three directions to avoid boundary effects in a finite-size system. Therefore, logical errors occur when the loops formed by combining error and correction chains span nontrivial $\mathbb{Z}_2$ cycles of the syndrome graph (that is a 3d torus in this case).

For our simulations, we consider a general error model for single-qubit and two-qubit gates. We also account for the photon loss and explain the decoding procedure later in this section.
Each single-qubit gate is accompanied by a 
single-qubit depolarizing channel
\begin{align}
{\cal D}_a (\rho)  = (1-p) \rho + \frac{p}{3} \sum_{P\in \{X,Y,Z\}} P_a \rho P_a,
\end{align}
with probability $p$. 
There are
several stages where this error may happen:
single-qubit gate on the emitter ($p_q$), and on photonic qubits ($p_S$), where we label various processes with the corresponding error rate.
Furthermore, we consider single-qubit errors on the emitter preceding a measurement ($p_T$) and
 single-qubit errors on photonic qubits preceding the final $X$-measurement ($p_F$) in the cluster state.
 
Similarly, a 
two-qubit gate is accompanied by a two-qubit depolarizing channel
\begin{align}
{\cal D}_{a,b} (\rho)  = (1-p') \rho + \frac{p'}{15} \sum_{\substack{P,P'\in \{I,X,Y,Z\}\\ (P,P')\neq (I,I)}} P_a P_b' \rho P_a P_b',
\end{align}
and we have two types of two-qubit gates: The emission process (c.f.~the $\cnot{q,n}$ gate in~(\ref{eq:emission})) with error probability $p_{2,q}$ and the $\cz{n,m}$ scattering gate with $p_2$.

As discussed in Sec.~\ref{sec:cluster states}, we generate linear clusters along $x$ axis and stitch them to form a 2d clusters in $xy$ plane which are then stitched to obtain a 3d structure. This way of generating a 3d resource state is clearly anisotropic and hence we have different effective error rates for the three kinds of qubits associated with the three faces ($xy$, $yz$, $xz$) of the primal (or dual) lattice. We further make use of the fact that at each instance of time the quantum state is a stabilizer state and we can transform errors to local errors up to stabilizer generators. This trick was also elucidated in Ref.~\cite{Wan2021}.
The overall single-qubit error rate for the qubits on $yz$ faces is given by
\begin{align}
\label{eq:qx}
q_x = \frac{4}{3} p_q + \frac{2}{3}( p_T + p_F) + \frac{8}{15} p_{2,q} + \frac{32}{15} p_{2},
\end{align}
while for the qubits on $xz$ and $xy$ faces, we have
\begin{align}
\label{eq:qyqz}
q_y = q_z = \frac{4}{3} p_q + \frac{2}{3}p_F + \frac{16}{15} (p_{2,q} + p_{2}),
\end{align}
where the subscript indicates the normal vector to the face on which the qubit lies.
Here, the error rates are calculated by noting that since qubits are measured in $X$ basis (as discussed earlier) then either $Z$ or $Y$ Pauli operators are considered a bit flip error (see Appendix~\ref{app: simulations} for details).

\begin{figure}
    \centering
    \includegraphics[scale=0.7]{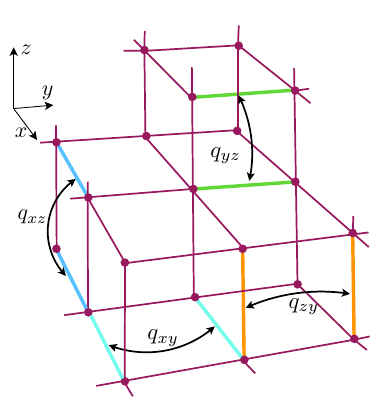}
    \caption{Correlated errors on the syndrome graph.}
    \label{fig:corr-error}
\end{figure}

Furthermore, two-qubit gates cause simultaneous errors on some qubits located on opposite edges of faces in syndrome graph,
\begin{align}
q_{yz} &= q_{zy} = \frac{4}{15} p_{2}, \nonumber \\
q_{xy} &=q_{xz} = \frac{4}{15} p_{2} + \frac{8}{15}  p_{2,q} ,
\label{eq:q-correlated}
\end{align}
where the subscript is read in the following way: The first and second letters denote the edge direction and the direction connecting two edges, respectively, as shown in Fig.~\ref{fig:corr-error}. We note that our state generation protocol implies that not all opposite edges experience a simultaneous error (see Appendix~\ref{app: simulations} for details).

\begin{figure}
    \centering
    \includegraphics[scale=0.63]{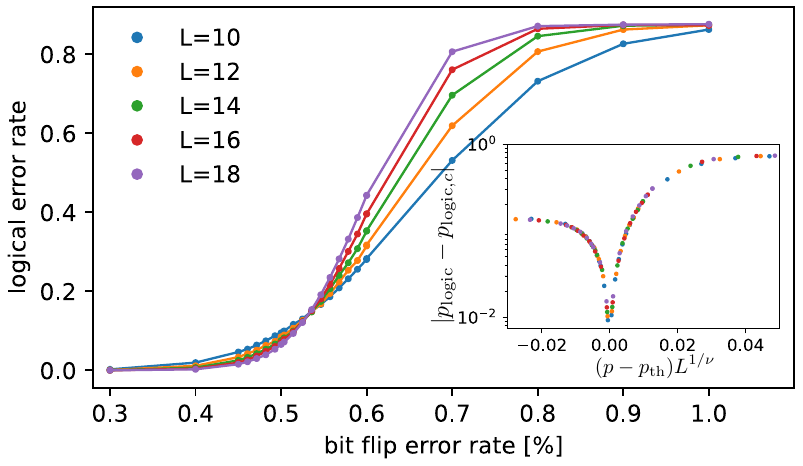}
    \caption{Error thresholds for the RHG state without loss. Ensemble size for averaging is $100$k. Inset shows the scaling collapse with $\nu=1$.}
    \label{fig:threshold}
\end{figure}

Using the above error model, we run Monte-Carlo simulations on finite-size lattices, and the results are plotted in Fig.~\ref{fig:threshold}. For simplicity, the error rates are set to be equal,
\begin{align}
	p_q=p_S=p_T=p_F=p_2=p_{2,q} = p.
\end{align}
Since the error model is anisotropic, the weights in the MWPM decoder are given in terms of their respective error probability $q_l$ as follows 
\begin{align}
\label{eq:weight}
    w_l = \log\left((1-q_l)/q_l\right),
\end{align}
and Dijkstra’s algorithm is exploited for finding shortest paths between every pair of vertices. We cross-checked our code with  PyMatching package~\cite{higgott2021pymatching}.
As shown in Fig.~\ref{fig:threshold}, we obtain the circuit error threshold of $p_\text{th} = (0.53\pm 0.02) \%$ which is close to the original theoretical model of Ref.~\cite{Raussendorf-ftcs}, that reported $0.58\%$.
Our circuit threshold can be further improved by better decoders which exploit sublattice correlations~\cite{Raussendorf_2007,PhysRevLett.98.190504} and take into account  degeneracies of possible matchings~\cite{Stace-fcts}. We postpone using decoders with either features to a future work.


\subsection{Loss error threshold}

In the remainder of this section, we discuss the effect of unheralded loss on the logical error rate. Loss in our case is detected during the measurement process and can be viewed as a quantum erasure channel where the error location is known but the error type is not known. The reason that here we are dealing with erasure (not deletion) is because every qubit is measured during the decoding process. Hence, the missing photons can be located through the detection process.  In terms of the construction process, if a qubit is lost at any point in our architecture, $\cz{}$ will not be active; however, the gate error rates on the present qubits remain the same.

Mathematically, loss error corresponds to partial tracing over the missing qubits. For instance, a loss (or erasure) channel for qubit $a$ is described by the following quantum channel,
\begin{align}
{\cal L}_a (\rho)  = (1-p_\text{loss}) \rho + p_\text{loss} \text{tr}_a(\rho) \otimes \ket{e}\!\bra{e},
\end{align}
where $\ket{e}$ denotes an unknown state outside the computational basis for qubit $a$, which in our case correspond to an empty (vacuum) state with no photon in either time bins. We should contrast this with presence/absence encoding~\cite{Wan2021} where the photon loss corresponds to a forced measurement (state collapse) to $\ket{0}$ rather than an erasure. 
To circumvent this issue, Ref.~\cite{Wan2021} proposed a dual-rail encoding (with two distinguishable photonic modes) by considering two additional energy levels in emitters. This idea however requires extra hardware overhead if one wants to measure the photonic qubits in bases other than the computational basis.

\begin{figure}
    \centering
    \includegraphics[scale=0.7]{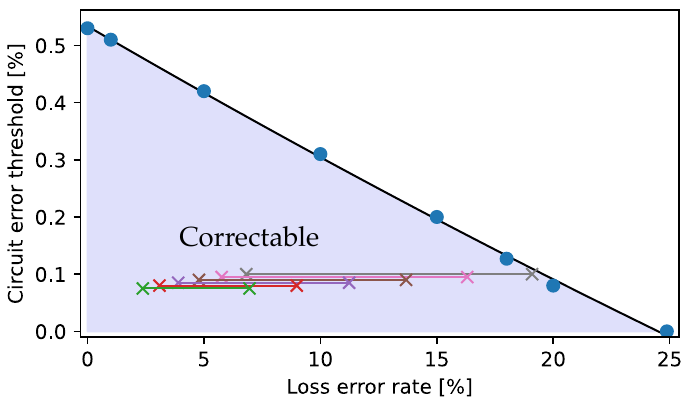}
    \caption{Error thresholds of the RHG state for various loss error rates and a quadratic fit (black solid line). Horizontal lines corresponds to curves (colors represent different sizes) in Fig.~\ref{fig:loss} where logical error rate is calculated. All lines are at $p=10^{-3}$ and displaced slightly vertically for clarity.}
    \label{fig:phasediag}
\end{figure}

One way to model the qubit loss is to reduce the stabilizer group to only stabilizer operators which commute with the Pauli operators acting on the missing qubit.  This effect leads to a syndrome graph on an irregular lattice~\cite{Stace-2d-pra,Stace-2d,Stace-fcts,Whiteside2014}. Suppose that a qubit connecting two nodes $c$ and $c'$ is lost (green edge in Fig.~\ref{fig:loss-syndrome}(c)). Since the two check operators associated with $c$ and $c'$, denoted by $S_c$ and $S_c'$, do not commute the lost qubit operators, we are only left with $S_c S_c'$ as a new stabilizer or supercheck operator. This effectively leads to a syndrome graph shown in Fig.~\ref{fig:loss-syndrome}(d)) which is no longer simple cubic. Now, in general multiple loss events occur and we end up with an irregular lattice with several supercheck vertices which may be connected to each other via more than one edge. For our decoder, we need to define a graph with only one edge connecting a pair of vertices. Suppose there are $n_i$ edges oriented along $i=x,y,z$ direction connecting two supercheck operators. We replace those edges with a single edge with an effective error rate
\begin{align}
\label{eq:eff_weight}
    p_l = \frac{1}{2}\left(1-(1-2q_{x})^{n_x}(1-2q_{y})^{n_y}(1-2q_{z})^{n_z}\right),
\end{align}
 which is simply derived by noting that syndrome defect occurs when odd number of those qubits are flipped. For the Monte-Carlo simulations we generate an error syndrome according to these modified error rates. The MWPM algorithm can be run on this graph, but finding whether the recovery chain causes a logical error or not is non-trivial since the graph is irregular, and it is not clear how to define $\mathbb{Z}_2$ cycles. Following Ref.~\cite{Stace-2d-pra}, we run the MWPM decoder on the original (cubic) lattice, where it is easy to determine the homology group of error recovery chains, provided that we put zero weights on missing edges (associated with loss error) and assign weights (\ref{eq:weight}) to the remaining edges according to Eq.~(\ref{eq:eff_weight}). For a given loss error rate, we calculate the logical error rate for several system sizes $L = 8,10,12,14,16$ (see Appendix~\ref{app: simulations} for simulation data), and find the error threshold. Figure~\ref{fig:phasediag} shows the circuit error threshold as a function of loss error probability $(p_\text{loss},p_\text{th})$, where we fit the data with a quadratic function and include the two limiting cases: No loss $(0,0.053)$ and when lost edges percolate $(0.249,0)$ and error correction always fails. The latter case corresponds to the bond percolation transition on a simple cubic lattice~\cite{Stace-fcts,Delfosse2020} (see Appendix~\ref{app:percolation}).
For reference, as we recall from Sec.~\ref{sec:cluster states} the RHG lattice can be constructed by measuring qubits on a body-centered lattice in a 3d cluster state. This method has several advantages such as simplicity and low hardware overhead; {however, the overall performance in terms of error thresholds is worse.} In Appendix~\ref{app:percolation}, we identify the loss threshold in this case to be a bond percolation transition on simple cubic lattice with first, second, and fourth nearest neighbors which occurs at $5.41\%$, which is significantly lower than directly generated RHG state. {We expect a similar behavior for the circuit error threshold, but do not further explore this numerically in this paper.}

\subsection{Delay-line trade off}

 Figure~\ref{fig:phasediag} may be interpreted as a 2d phase diagram in the sense that if the system characteristics $(p_\text{loss},p)$ fall within the shaded area (correctable) then we are in the fault-tolerant regime, i.e.~the logical error rate can be reduced exponentially as we increase the cluster dimension $L$. 
 However, in practice as we increase the cluster size, the dominant factor to the photon loss will be the optical fiber where the cumulative loss probability grows with system size as in
 \begin{align}
 \label{eq:fiber-loss}
 p_\text{loss} = 1- 10^{-\frac{\lambda}{10} v\tau L^2 },
 \end{align}
 since we are using an optical fiber delay line to store $L^2$ qubits. Here, $\lambda$ denotes the loss per unit length (in units of dB/km), $v = 2.13\times 10^{8}$ m/s  is the speed of light in the fiber and $\tau$ is the time difference between two consecutive photonic qubits as discussed in Sec.~\ref{sec:design} (c.f.~Fig.~\ref{fig:emitter}(b)), which varies over the range of $\tau \approx 0.1$-$50$nsec~\cite{Uppu2021}. Therefore, enlarging the cluster state size eventually overwhelms the error correction and takes us outside the correctable region of the phase diagram.


\begin{figure}
    \centering
    \includegraphics[scale=0.7]{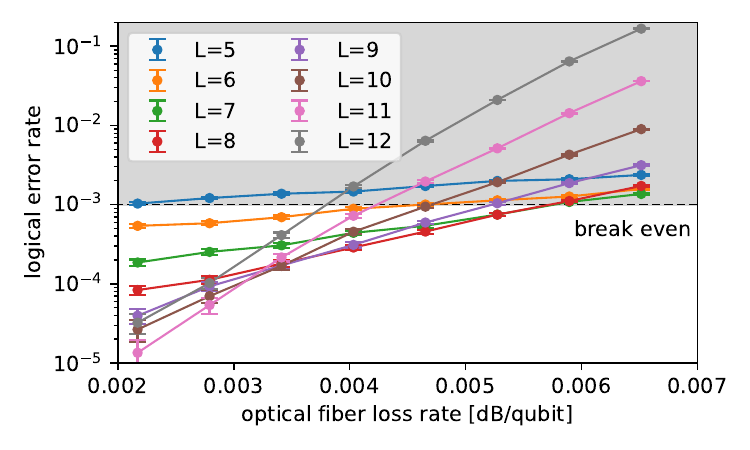}
    \caption{Logical error rate for different system sizes as a function of optical fiber loss rate per qubit, i.e.~$\lambda v \tau$ defined in Eq.~(\ref{eq:fiber-loss}). Here, we assume physical qubit error rates are $p=0.1\%$. The ensemble average for each point is performed on $N_s = 5\times 10^5$ samples and error bars $\sqrt{p_\text{fail}(1-p_\text{fail})/N_s}$ are shown when applicable. Solid lines are guide for eyes.}
    \label{fig:loss}
\end{figure}

Given that the loss rate increases with the system size, one may ask whether this setup can be exploited to reduce the logical error rate and eventually be used as a fault-tolerant quantum memory. To address this question, let us run a numerical experiment: We start with a physical error rate $p=p_0$ and ask whether a typical loss rate in optical fiber can allow for smaller logical error rates upon going to larger cluster states. That is to ask, does error correction in our scheme work given the typical values in real systems? We set $p_0 = 10^{-3}$ and compute the logical error rate as a function of loss per length in number of qubits [dB/qubit] for various system sizes as plotted in Fig.~\ref{fig:loss}. As we see, if the loss rate obeys $\lambda v \tau < 5.7\times 10^{-3}$ dB/qubit, the logical error rate {can be made} smaller than the circuit error rate {for some values of $L$}. Typical fiber losses are $\lambda = 3.5$ dB/km at the native QD operation wavelength ($950$ nm)~\cite{Uppu2021}, and improve to $\lambda = 0.2$ dB/km at the telecom C-band. Hence, for the protocol to work without a frequency converter, the emission rate needs to be $\tau \lesssim 7.6$ nsec so that the break-even point for the error rate occurs at $3.5$ dB/km.

 As mentioned, for a fixed circuit error rate as we increase the cluster size, the loss error rate also increases. For instance, points in Fig.~\ref{fig:loss} for various loss rates per qubit and system sizes correspond to spanning horizontal lines in the phase diagram as shown in. Fig.~\ref{fig:phasediag}. In other words, as we increase the system size, the new system corresponds to a point further right in the phase diagram. We should note that at each loss rate per qubit the logical error rate is bounded from below, and there is an optimum system size  $L_\text{opt}$ which gives the lowest possible logical error rate. Ref.~\cite{Wan2021} showed that the optimal logical error rate associated with  $L_\text{opt}$ follows an empirical formula $p_\text{opt}\sim \exp(-c' \lambda^{-1/2})$. In Appendix~\ref{app:Lopt}, we use the saddle point approximation and derive this relation by assuming an exponentially decaying (in cluster dimension) logical error rate within the correctable region of the phase diagram. However, our numerical data is not conclusive in the sense that a wide range of exponents $0<x<1$ in $p_\text{opt}\sim \exp(-c' \lambda^{-x})$ can fit our data well. We suspect the data in Ref.~\cite{Wan2021} may suffer from the same issue. We attribute this observation to the finite-size effects and possible invalidity of the assumption that the logical error rate decays exponentially with the system size. 


Last but not least, in all finite-size scaling analyses (with or without loss), we estimate the circuit error threshold and the critical exponent by collapsing the data on a single-parameter scaling ansatz
\begin{align}
	p_{\text{logic}} (p,L) - p_{\text{logic}}(p_\text{th},L) = f((p-p_\text{th}) L ^{1/\nu}),
\end{align}
without a polynomial fit to the data following the method developed in Ref.~\cite{Skinner2019} (see details in Appendix~\ref{app: simulations}),
where $p_{\text{logic},c}=p_{\text{logic}}(p_\text{th},L)$ denotes the logical error rate at the scale-invariant (critical) point.
We find that  $\nu =  1.00 \pm 0.05$ (e.g., see inset of Fig.~\ref{fig:threshold}) which matches the critical exponent of the transition in the random plaquette $\mathbb{Z}_2$ gauge theory in 3D~\cite{WANG200331}. We also observe that $p_{\text{logic},c}$ in our case is close to those of Refs.~\cite{WANG200331,Raussendorf-ftcs,Stace-fcts}.
This further indicates that our numerics (despite detailed differences with previous works) probe the same phase transition since the value $p_{\text{logic},c}$ must be related to some universal physical observable such as the conductivity which only depends on the universality class. 



Let us conclude this section by reiterating a general remark about fault-tolerant photonic quantum memories based on discrete-variable cluster states on RHG lattice. As we discussed earlier in this section, there is a hard upper bound on loss tolerance in such systems (regardless of the protocol) corresponding to a bond percolation transition in a simple cubic lattice which occurs at {$p_\text{perc.} = 24.9\%$}. The critical value in turn translates into $1.24$ dB transmission loss, which imposes an upper bound on the allowed overall loss. In other words, quantum error correction in such schemes would not work if the total transmission loss is greater than $1.24$ dB, regardless of how low the qubit (bit flip) error rate can be made. Assuming that the loss is dominated by the optical fiber, it is easy to show that the largest fault-tolerant cluster is given by,
\begin{align}
L_\text{max} =  \left|\frac{10}{\lambda v\tau} \log_{10} (1-p_\text{perc.})\right|^{1/2}.
\end{align}
Plugging in $\lambda =0.2$ dB/km, we find that $L_\text{max}$ ranges between  $24$ to $540$ corresponding to $\tau = 50 $ nsec and $\tau = 0.1 $ nsec, respectively. An interesting observation here is that if the emission rate is fast enough $\tau^{-1}> 1/3$ GHz the photon loss in the delay line does not fundamentally preclude the realization of a quantum memory with million qubits.

\section{Discussion}
\label{sec:conclusions}


 In conclusion, we propose a new optical circuit design based on quantum emitters and optical fiber delay lines to deterministically generate resource graph states.
Our design consists of an active component and a stack of passive ones which are responsible for single photon generation and gating, respectively. As we show, different graph states can be obtained by varying the resonant pulse sequence applied to the quantum emitter and possibly additional single-qubit gates. Our optical circuit admits a modular structure, where passive components can be stacked to produce graphs on higher dimensional lattices.
We provide explicit design protocols for generating cluster states in any dimension, repeater states, and tree graph states. We further discuss an error model for our design and calculate the circuit error threshold in the case of fault-tolerant cluster state. We map out an error-correction phase diagram in terms of circuit error and loss rates in our setup and identify the correctable region. In the absence of loss, we find the circuit error threshold to be $0.53\%$, while in the absence of circuit error, the loss threshold is $24.9\%$ corresponding to bond percolation on a cubic lattice. Finally, we show through an example that the break-even point for error reduction in the fault-tolerant construction can be reached with the existing technology. 

There have been several recent studies on using quantum emitters to deterministically generate graph states for quantum communication~\cite{Economou2017,Russo_2019,Hilaire2021resource,Li2021,PhysRevLett.116.093601}
as well as quantum computing~\cite{Pichler2017,Wan2021}. 
A central theme in all these works (including ours) is to reduce the number of components by optimizing the generation protocol. This is particularly beneficial in terms of reducing the engineering challenges to achieve the desired outcome,  as in such systems improving a small number of components leads to a great performance enhancement.
Nonetheless, these minimal architectures pose new challenges.
In this work, we address some of these issues in a more realistic design. Our circuit contains only one actively monitored emitter and does not require any two-qubit gates for emitters. Having one emitter implies a reference clock cycle which is governed by the resonant pulse sequence applied to the emitter. Our protocol is based on time-bin qubit encoding which is not sensitive to dephasing error and equipped with arbitrary single-qubit gate set for performing computation or other manipulations on  the output resource state.
Furthermore, there is no need for ultrafast optical switches in our design.


At the hardware level, our proposed device architectures are inspired by the platforms based on semiconductor QDs coupled to a waveguide, although they are not restricted to such platforms and may be applicable to other quantum emitter technologies such as atomic systems
where atom-photon entanglement and quantum memory were demonstrated~\cite{Togan2010,Duan2001}. As mentioned above, there is a hard upper bound for loss error threshold depending on the generation protocol of the RHG lattice:
We report $24.9\%$ for direct generation and $5.41\%$ for indirect generation by measuring 3d cluster state,
which translate into $1.24$ and $0.24$ dB transmission loss, respectively. Reaching the levels below the loss threshold requires minimizing the signal loss at several locations in a photonic system most notably at the interfaces between different components and inside the optical fiber delay line. As we discussed in Sec.~\ref{sec:error}, one should be able to deal with the loss in the optical fiber as long as the photon emission rate is fast enough, which is currently achievable~\cite{Uppu2021}.
Another place for improvement is the QD coupling to a single mode waveguide or nanocavity which needs to be strong enough while coupling to unwanted leaky modes must be suppressed; this is quantified by the single-photon coupling efficiency or $\beta$ factor (i.e., the ratio of the number of emitted photons in the guided mode to that of all modes), which routinely reaches near unity in nanophotonic devices. See for example Refs.~\cite{Arcari2014,Uppu2021} which reports $\beta>98.4\%$ implying a quantum cooperativity (i.e., the ratio between the number of emitted photons in the guided mode and that of undesired modes) of $C=\beta/(1-\beta)\geq 62$. 
Photon indistinguishability is also an essential requirement for discrete-variable quantum information processing 
which has been achieved for single QD~\cite{Senellart2016}; however, having multiple QDs emitting nearly indistinguishable photons is quite challenging. Here, we tried to minimize that by having only one active device as a photon source and use passive devices containing multiple QDs which reduce the chance of incompatibility.

We close the discussion with a survey of new avenues for future research.
In this paper, we base our design on quantum emitters with a $\Lambda$-type energy level-structure and use them to generate linear cluster states as part the desired graph states. However, the three energy levels are not so essential in our design, and a two-level emitter can work as well. As mentioned in Sec.~\ref{sec:design}, one way to generate a linear cluster state with a two-level emitter is by using the circuit in Fig.~\ref{fig:repeater-schematic} and putting an MZI with a delay line as in Eq.~(\ref{eq:MZI}) which effectively generates a sequence of time-bin qubits in $\ket{+}$. The drawback is that we have one additional $\cz{}$ layer in our passive component stack.  It would be interesting to fully examine the error model in this case. Upon first glance, it looks like the error model is identical to that of the original work of Raussendorf \emph{et~al.}~\cite{Raussendorf-ftcs} where the error threshold is $0.58\%$. Here, we explained how to generate some prototypical graph states; an interesting direction would be to develop a general algorithm to produce arbitrary graphs using emitter with delay lines, perhaps along the lines of Ref.~\cite{Li2021}.

Throughout this work, we use a standard MWPM decoder to calculate the circuit error thresholds for the fault-tolerant cluster state on the RHG lattice. In principle, the thresholds can be further improved by using a better decoder.
As mentioned earlier, cluster states on the RHG lattice can tolerate a maximum loss error rate of $24.9\%$. As shown in Refs.~\cite{Nickerson2018,Newman2020}, this threshold can be improved significantly to larger values (near $50\%$) by realizing graph states on more complex (beyond foliated) 3d crystal structures. Designing protocols to generate such states could be worth pursuing.  Here, we focus on the state generation and discuss the fault-tolerant properties of the 3d cluster state as a quantum memory. A natural next step is to define protocols for performing logical gates on these states and envision a multi-qubit setup with a fault-tolerant logical gate set.
One of the building blocks of our design was the $\cz{}$ scattering block~\cite{Jeannic2021,Schrinski2021} based on array of QDs. The nonlinearity induced by this device could enable a new capability for non-Gaussian operations in continuous-variable quantum information processing schemes possibly via a hybrid discrete-continuous variable protocol.



\acknowledgements

The authors acknowledge insightful discussions with Bing Qi, Peter Lodahl, Ramana Kompella, and Stephen DiAadamo. We especially thank Galan Moody, Stefano Paesani, and Isaac Kim for valuable discussions and helpful comments on the manuscript.

\appendix

\section{Details of error threshold simulations}
\label{app: simulations}

\renewcommand\theequation{A\arabic{equation}}

In this appendix, we show how to derive the error rates used in the main text, Eqs.~(\ref{eq:qx}), (\ref{eq:qyqz}) and (\ref{eq:q-correlated}). We also provide some raw data for the phase diagram and briefly discuss how the error threshold and critical exponent are computed.

\begin{figure*}
    \centering
    \includegraphics[scale=0.78]{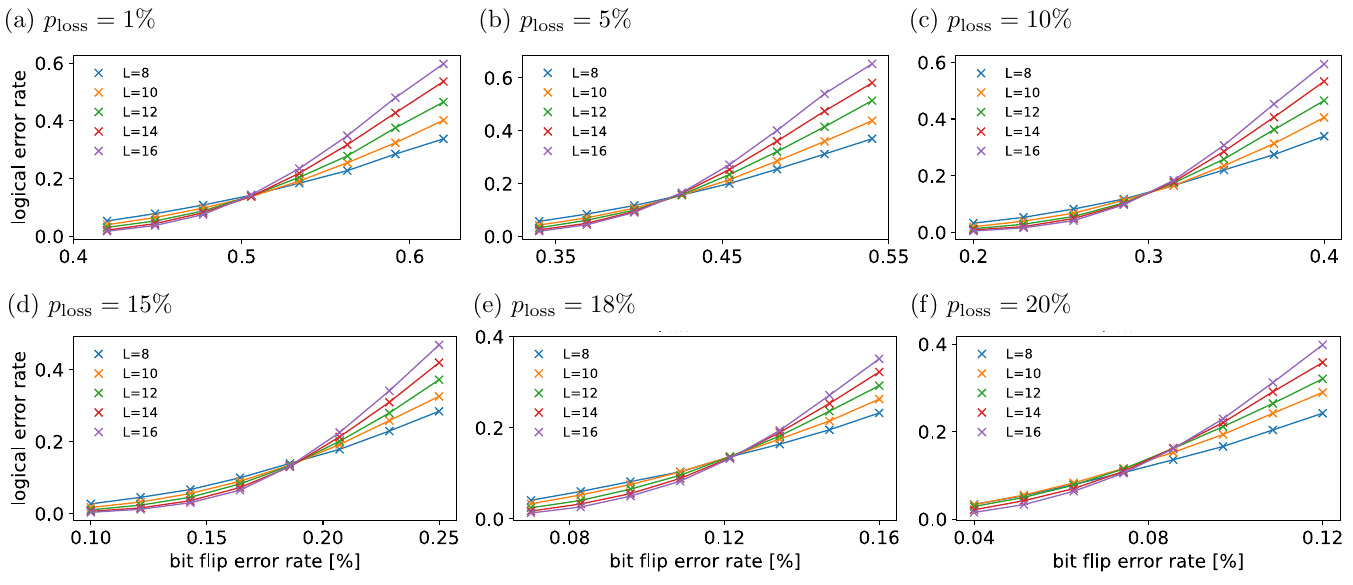}
    \caption{Error thresholds of the RHG state for various loss rates. Ensemble size for averaging is $60$k.}
    \label{fig:loss-threshold-plots}
\end{figure*}

\subsection{Effective circuit error rates}

In this part, we derive the effective circuit error rates for the cluster state on RHG lattice given the error processes in our setup. We only consider lowest order contribution in probability amplitudes as we are only exploring small error regimes $\lesssim 10^{-2}$.

Let us begin with qubits on the $yz$ faces (normal to the $x$ axis). They can be generated via the process in Eq.~(\ref{eq:photon-hadamard}) where the initial state of the emitter is $\ket{0}$. After the first hadamard gate 
or before the  final qubit measurement, $Z_q$ or $Y_q$ error gives $Z_n$ error with probabilities $\frac{2}{3}p_q$ and $\frac{2}{3}p_F$, respectively. Also, after the second hadamard gate or before the emitter measurement, $X_q$ or $Y_q$ error leads to a $Z_n$ error on the qubit  with probabilities $\frac{2}{3}p_q$ and $\frac{2}{3}p_T$, respectively. Note that we considered errors up to instantaneous stabilizers. Regarding two-qubit gates: first, $\cnot{q,n}$ gives  $Z_n$ via the errors $(Z_q, Y_q) \otimes (I_n,X_n)$ 
and
$(I_q,X_q) \otimes (Z_n,Y_n)$ 
with overall probability $2\times\frac{4}{15} p_{2,q}$. Second, $x$-type qubit undergoes four $\cz{}$ gates where $(Y_n, Z_n) \otimes P_{n'}$ Kraus operators give a circuit error with probability $4\times \frac{8}{15} p_{2}$.
Hence, we obtain
\begin{align}
q_x = \frac{4}{3} p_q + \frac{2}{3}( p_T + p_F) + \frac{8}{15} p_{2,q} + \frac{32}{15} p_{2}.
\end{align}

The qubits on the $xz$ and $xy$ faces (or $y$ and $z$ type qubits, respectively) are part of the linear cluster generated by the emitter $Q$, where the emitter error can be transformed into a qubit error by using an instantaneous stabilizer at each moment, i.e., $X_q \otimes Z_n$ after emission of $n$-th qubit. At every qubit emission process, we effectively apply the operator 
\begin{align}
A_n =  H_q \cnot{q,n}.
\end{align}
As far as $H_q$ gate is concerned, a $Z_q$ error before $A_n$ and an $X_q$ after $A_n$ lead to a $Z_n$ error with probability of $\frac{2}{3}p_q$. $Y_q$ error before $A_n$ causes a two-qubit $Z_n Z_{n-1}$ error with prob $p_q/3$. Since qubits $n$ and $n-1$ belong to primal and dual lattices, then this error is simplified into  single-qubit errors. Plus, qubits experience $Z$ or $Y$ error right before measurement with probability $\frac{2}{3}p_F$. As for two qubit gates: Similar to $x$-type qubits $\cnot{q,n}$ leads to $Z_n$ with probability $2\times\frac{4}{15} p_{2,q}$.
Second, $\cnot{q,n-1}$ gives $Z_n$ through 8 Kraus terms  $(X_q, Y_q) \otimes P_{n-1}$ with probability $\frac{8}{15} p_{2,q}$. Third, these types of qubits experience two $\cz{}$ gates where $(Y_n, Z_n) \otimes P_{n'}$ Kraus operators are  circuit errors with probability $2\times \frac{8}{15} p_{2}$. We note that there is no error propagation from neighboring qubits since they are prepared in the hadamard basis (i.e., eigenstates of $X$ operator). Hence, the overall single-qubit error rates for $y$ and $z$ type qubits is given by
\begin{align}
q_y = q_z = \frac{4}{3} p_q + \frac{2}{3}p_F + \frac{16}{15} (p_{2,q} + p_{2}).
\end{align}








Next, we consider correlated errors. Let us reemphasize that nearest neighbor qubits on the RHG lattice belong to primal and dual lattice and next nearest neighbor qubits belong to the same parity check. Hence, they are not considered correlated errors. In our setup, we only generate correlated errors between opposite edges of  each face on the primal lattice or neighboring faces on the same plane (which correspond to opposite edges of faces on the syndrome graph (c.f.~Fig.~\ref{fig:corr-error})).
For all types of qubits, we may consider two $\cz{}$ gates connecting two next nearest neighbor qubits, where 
$(X_k, Y_k) \otimes (Y_{n}, Z_{n})$ error after the first gate ($k$ denotes the shared qubit for the two $\cz{}$ gates) leads to a correlated error with probability $\frac{4}{15} p_{2}$. For $x$-type qubits, we have an additional error due to {$P_q \otimes (X_k, Y_k)$} error with probability $\frac{8}{15}  p_{2,q}$ from {$\cnot{q,k}$} gate during the linear cluster generation process. Therefore, we may write
\begin{align}
q_{yz} &= q_{zy} = \frac{4}{15} p_{2}, \\
q_{xy} &=q_{xz} = \frac{4}{15} p_{2} + \frac{8}{15}  p_{2,q}.
\end{align}


\subsection{Simulation data for the phase diagram}

As discussed in Sec.~\ref{sec:error}, we run Monte-Carlo simulations to find the circuit error threshold for various loss rates and map out a phase diagram for our design. The simulation results are summarized in Fig.~\ref{fig:loss-threshold-plots}.
We use the algorithm of Ref.~\cite{Skinner2019} for estimating the circuit error threshold $p_\text{th}$ and the correlation length exponent $\nu$ which is based on searching for scaling collapse among curves $p_\text{logic}(p, L) - p_\text{logic}(p_\text{th}, L)$ as a function of the single variable $(p-p_\text{th})L^{1/\nu}$. Below, we briefly discuss this method.

The objective is to minimize a cost function $R(p_\text{th}, \nu)$ which 
essentially measures deviation from a data collapse on a universal (unknown) function.
The corresponding optimal values are our estimates for $p_\text{th}$ and $\nu$. For a given value of $p_\text{th}$ and $\nu$, we estimate $p_\text{logic}(p_\text{th}, L)$ for each system size $L$ by a piece-wise linear interpolation. We then calculate the function $y_L (x)= p_\text{logic}(p, L) - p_\text{logic}(p_\text{th}, L)$ and  $x=(p-p_\text{th})L^{1/\nu}$ for given values of $p$ and $L$ from the dataset. This gives a family of curves $y_L(x)$ vs.~$x$, which we wish to collapse on a single curve. Next, we sample from these curves at a set of discrete points $x_i$ and define the cost function
\begin{align}
R = \sum_{i, L} \left[ y_L(x_i) - \bar{y}(x_i) \right]^2,
\end{align}
in terms sum of the variance of $y_L(x_i)$ for different system sizes, where \begin{align}
    \bar{y}(x_i) = \frac{1}{N_L} \sum_L y_L(x_i),
\end{align}
is the mean value at point $x_i$, and $N_L$ is the number of different system sizes in the dataset. We should note that again we use piece-wise linear interpolation to estimate $y_L(x_i)$ and omit a point for a given $L$ if it is outside the range of the dataset for that particular $L$.
Finally,  we search numerically for the values of $p_\text{th}$ and $\nu$ that minimize the objective function.

In order to estimate the uncertainty in our results for $p_\text{th}$ and $\nu$, we examine how the optimum point change when we run the minimization algorithm on a subset of data. In particular, we choose every pair of system sizes and calculate the optimum values of $p_\text{th}$ and $\nu$ and use their variation as a proxy for the uncertainty.

\section{Loss error threshold and percolation transition}
\label{app:percolation}

As mentioned in the main text, there is a loss threshold for a fault-tolerant graph state similar to the circuit error threshold above which the quantum error correction is not effective. Their difference is that the location of loss errors are known.
In this appendix, we discuss the connection between loss error and bond percolation transition where the critical probability in the percolation problem correspond to the loss threshold.

Consider a syndrome graph where the qubits and parity checks are associated with edges and vertices of the graph. For instance, syndrome graph for the RHG cluster state forms a simple cubic lattice (Fig.~\ref{fig:loss-syndrome}). Loss error is modeled as independent random processes on every edge with probability $p$. When a qubit is lost (erased), the parity checks on the adjacent vertices are not well-defined. Instead, one can define supercheck operators which are product of even number of the adjacent parity checks, so that the new stabilizers do not have a Pauli operator acting on the lost qubit. For a given set of lost qubits, we check for a logical error event by determining whether the error chain form a non-contractible loop or not. This is exactly the bond percolation problem on the syndrome graph. This means that below the critical probability $p_\text{perc.}$ in percolation problem increasing the system size decreases the probability of forming a percolating path. In the quantum error correction language, this is equivalent to saying that below the loss threshold $p_\text{th}$ the logical error probability can be arbitrarily reduced by increasing the system size. In other words, $p_\text{th}=p_\text{perc.}$.
So, the loss threshold for the RHG lattice is described by the bond percolation on the simple cubic lattice and is equal to $p_\text{perc.} \simeq 24.9\%$.

Now, let us consider another example when we start with a 3d cluster state (i.e., simple cubic lattice) and remove a subset of qubits on the body-centered lattice in a doubled unit cell, i.e., the center of cell and its six corners (c.f.~Fig.~\ref{fig:clusterstate}(b)), by measuring them in $Z$ basis.
Before the measurement, the product of stabilizers on faces of a cell $c$ is given by
\begin{align}
    S_c = \bigotimes_{f\in \partial c} S_f= \bigotimes_{i} Z_{c\pm 2 \textbf{e}_i} \bigotimes_{f\in \partial c} X_f.
\end{align}
Notice the extra operators in front of the actual stabilizer in Eq.~(\ref{eq:cube}).
Here, $\textbf{e}_i$ with $i=x,y,z$ denotes the basis vectors in the cluster state cubic lattice. So, $c\pm 2 \textbf{e}_i$ correspond to six fourth nearest neighbor qubits (distance $2a$ with lattice constant $a$) of the center of the cell.
If there is no loss, measuring these qubits simply replace them with $\pm$ depending on the measurement outcome. However, if these qubits are lost, we need to combine those stabilizers sharing them. Same applies independently to the stabilizers associated with the dual lattice.
We note that each of the qubits to be removed is shared between six parity checks located at $\pm 2 \text{e}_i$ with respect to that qubit.
Hence, given a missing qubit with probability $p$, we must connect the parity check on cell $c$ with its second (distance $2\sqrt{2} a$) and fourth (distance $4 a$) nearest neighbors on the syndrome graph. Therefore, threshold in this case corresponds to the bond percolation transition on cubic lattice with not only nearest neighbor links but also links connecting to the second and fourth nearest neighbors. 
To estimate the loss error threshold, we use the formula $p_\text{perc.}\sim \frac{1}{z-1}$ where $z$ is total coordination number of a lattice point~\cite{PhysRevE.53.2177}. Having known $p_\text{perc.}\simeq 24.9\%$ for simple cubic lattice where $z=6$, we find that $p_\text{perc.}\simeq 5.41\%$ since $z= 6+ 12 + 6 $.

\section{Optimum logical error rate with fiber delay lines}
\label{app:Lopt}

\renewcommand\theequation{B\arabic{equation}}

In the main text, we learn that increasing the cluster size does not necessarily lead to decreasing logical error rate since the cumulative loss rate grows due to the signal propagation through the delay lines. For a given loss error rate per qubit in the fiber, $\eta$, we have the overall loss as $p_\text{loss} = 1-e^{-\eta L^2}$, where $\eta$ is related to $\lambda$ in Eq.~(\ref{eq:fiber-loss}) via $\eta = \lambda v \tau \ln(10) / 10$ [because of using a different base]. 
In this appendix, we use  the saddle point approximation and show that there is an optimum value for $L$ where the logical error rate is minimum and obeys the form $p_\text{opt}\sim\exp(-c'\eta^{-1/2})$. 

\begin{figure}
    \centering
    \includegraphics[scale=0.63]{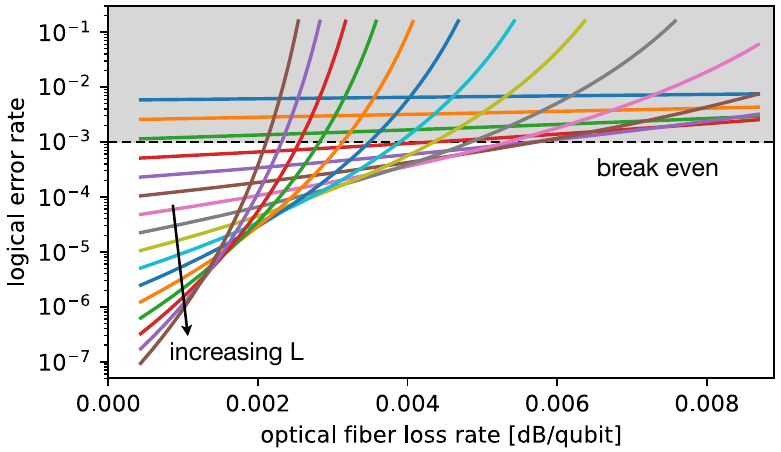}
    \caption{Logical error rate for different system sizes using exponentially decaying logical error rate in the correctable region of the phase diagram, e.g.~Eq.~(\ref{eq:exp_decay_logical}). The system size is varied from $L=4$ to $20$ as shown by the arrow.}
    \label{fig:loss-simulated}
\end{figure}
We start by assuming an exponential form for the logical error rate,
\begin{align}
\label{eq:exp_decay_logical}
p_\text{logic} &\sim \left( \frac{p}{p_\text{th}(p_\text{loss})} \right)^{\frac{L}{2}} \\
&=  \left( \frac{p}{\alpha \eta L^2 + \beta} \right)^{\frac{L}{2}} \\
&=  e^{f(L)} 
\end{align}
where we use a linear approximation to the error threshold as a function of loss rate (phase boundary in Fig.~\ref{fig:phasediag}),
\begin{align}
p_\text{th}(p_\text{loss}) &= \alpha p_\text{loss} + \beta  \nonumber \\
&=  \alpha (1-e^{-\eta L^2}) + \beta \nonumber \\
&\approx \alpha \eta L^2 + \beta,
\end{align}
and define
\begin{align}
	f(L) =\frac{L}{2} ( \ln(p)  - \ln(\alpha \eta L^2 + \beta) ).
\end{align}
The ansatz (\ref{eq:exp_decay_logical}) is plotted in Fig.~\ref{fig:loss-simulated} for various system sizes where we use the fit to the phase boundary in Fig.~\ref{fig:phasediag} for $p_\text{th}(p_\text{loss})$. We observe that the analytical curves are qualitatively similar to the numerical results in Fig.~\ref{fig:loss}. However, we could not find a good quantitative fit to our data using this ansatz. 
To find the minimum logical error, 
we use the saddle point approximation by setting $f'(L_\text{opt}) = 0$, and we
get
\begin{align}
	( \ln(p)  - \ln(\alpha \eta L_\text{opt}^2 + \beta))  - \frac{2\alpha \eta L_\text{opt}^2}{(\alpha \eta L_\text{opt}^2 + \beta)} =0.
\end{align}
The above equation can be turned into an algebraic equation independent of $\eta$ for an axiliary variable $x_0 = \alpha \eta L_\text{opt}^2$. Hence, we can plug in the solution $L_\text{opt}= \sqrt{x_0/\eta \alpha}$ to Eq.~(\ref{eq:exp_decay_logical})
and obtain
\begin{align}
p_\text{logic}  &\approx  \left( \frac{p}{x_0+ \beta} \right)^{\frac{1}{2}\left(\frac{x_0}{\eta \alpha}\right)^{1/2}} 
\end{align}
which is exponentially decaying in $\eta^{-1/2}$.

\bibliography{refs.bib}

\end{document}